\documentclass[journal]{IEEEtran}
\usepackage{epsf,exscale,times}
\usepackage{graphicx}
\usepackage{amssymb}
\usepackage{mathtools}
\usepackage{subcaption}
\usepackage{multirow}
\usepackage{multicol}
\usepackage{glossaries}
\usepackage{cite}
\usepackage{nccmath}
\usepackage{commath}
\usepackage{comment}
\usepackage{color}
\usepackage{algorithm2e}
\usepackage{makecell}
\usepackage{bm}

\usepackage{array}
\newcolumntype{P}[1]{>{\centering\arraybackslash}p{#1}}
\newcolumntype{M}[1]{>{\centering\arraybackslash}m{#1}}

\newacronym{FNR}{FNR}{Luxembourg National Research Fund}
\newacronym{SNR}{SNR}{Signal to Noise Ratio}
\newacronym{INR}{INR}{Interfernece to Noise Ratio}
\newacronym{SINR}{SINR}{Signal to Interference plus Noise Ratio}
\newacronym{AF}{AF}{Ambiguity Function}
\newacronym{MIMO}{MIMO}{Multiple-Input Multiple-Output}
\newacronym{SISO}{SISO}{Single-Input Single-Output}
\newacronym{SIMO}{SIMO}{Single-Input Multiple-Output}
\newacronym{CD}{CD}{Coordinate Descent}
\newacronym{BCD}{BCD}{Block Coordinate Descent}
\newacronym{GD}{GD}{Gradient Descent}
\newacronym{MM}{MM}{Majorization-Minimization}
\newacronym{FMCW}{FMCW}{Frequency Modulated Continuous Wave}
\newacronym{PMCW}{PMCW}{Phase Modulated Continuous Wave}
\newacronym{DFT}{DFT}{Discrete Fourier Transform}
\newacronym{FFT}{FFT}{Fast Fourier Transform}
\newacronym{MVDR}{MVDR}{Minimum Variance Distortionless Response}
\newacronym{MBI}{MBI}{Maximum Block Improvement}
\newacronym{RFPA}{RFPA}{Radio Frequency Power Amplifier}
\newacronym{BPSK}{BPSK}{Binary Phase Shift Keying}
\newacronym{QPSK}{QPSK}{Quadrature Phase Shift Keying}
\newacronym{ULA}{ULA}{Uniform Linear Array}
\newacronym{DOF}{DOF}{Degrees of Freedom}
\newacronym{PSK}{PSK}{Phase Shift Keying}
\newacronym{PSL}{PSL}{Peak Sidelobe Level}
\newacronym{PSLR}{PSLR}{Peak Sidelobe Level Ratio}
\newacronym{ISL}{ISL}{Integrated Sidelobe Level}
\newacronym{ISLR}{ISLR}{Integrated Sidelobe Level Ratio}
\newacronym{LFM}{LFM}{Linear Frequency Modulation}
\newacronym{CPI}{CPI}{Coherent Pulse Interval}
\newacronym{RCS}{RCS}{Radar Cross Section}
\newacronym{CNR}{CNR}{Clutter to Noise Ratio}
\newacronym{MTI}{MTI}{Moving Target Indicator}
\newacronym{ROC}{ROC}{Receiver Operating Characteristic}
\newacronym{MPSK}{MPSK}{$M$-ary Phase Shift Keying}
\newacronym{PAR}{PAR}{Peak-to-Average Ratio}
\newacronym{GFP}{GFP}{Generalized Fractional Programming}
\newacronym{PRI}{PRI}{Pulse Repetition Interval}
\newacronym{PRF}{PRF}{Pulse Repetition Frequency}
\newacronym{MMM}{MM}{Majorization Minimization or Minorization Maximization}
\newacronym{QCQP}{QCQP}{Quadratic Constraint Quadratic Programming}
\newacronym{SDP}{SDP}{Semi-definite Programming}
\newacronym{CCL}{CCL}{Cross-Correlation Level}
\newacronym{TDMA}{TDMA}{Time-Division Multiple Access}
\newacronym{FDMA}{FDMA}{Frequency-Division Multiple Access}
\newacronym{CDMA}{CDMA}{Code-Division Multiple Access}
\newacronym{DDMA}{DDMA}{Doppler-Division Multiple Access}
\newacronym{TDM}{TDM}{Time Division Multiplexing}
\newacronym{FDM}{FDM}{Frequency Division Multiplexing}
\newacronym{CDM}{CDM}{Code Division Multiplexing}
\newacronym{DDM}{DDM}{Doppler Division Multiplexing}
\newacronym{SDR}{SDR}{Semi-definite Relaxation}
\newacronym{QSDR}{QSDR}{Quantized Semi-definite Relaxation}
\newacronym{CA}{CA}{Cyclic Algorithm}
\newacronym{ADMM}{ADMM}{Alternating Direction Method of Multipliers}
\newacronym{PDR}{PDR}{Projection, Descent, and Retraction}
\newacronym{SQP}{SQP}{Semidefinite Quadratic Programming}
\newacronym{CM}{CM}{Constant Modulus}
\newacronym{MPS}{MPS}{Minimum Peak Sidelobe}
\newacronym{BiST}{BiST}{Binary Sequences seTs}
\newacronym{ESA}{ESA}{Effective Simulated Annealing}
\newacronym{BSUM}{BSUM}{Block Successive Upper Bound Minimization}
\newacronym{CS}{CS}{Compressive Sensing}
\newacronym{CAN}{CAN}{Cyclic Algorithm-New}
\newacronym{CW}{CW}{Continuous Wave}
\newacronym{WeBEST}{WeBEST}{Weighted BSUM sEquence SeT}
\newacronym{MISL}{MISL}{Monotonic minimizer for Integrated Sidelobe Level}
\newacronym{DP}{DP}{Discrete Phase}
\newacronym{CP}{CP}{Continuous Phase}
\newacronym{SAR}{SAR}{Synthetic-Aperture Radar}
\newacronym{CFAR}{CFAR}{Constant False Alarm Rate}

\newcommand{\bA}{\mbox{\boldmath{$A$}}}
\newcommand{\ba}{\mbox{\boldmath{$a$}}}

\newcommand{\bX}{\mbox{\boldmath{$X$}}}

\newcommand{\bzero}{\mbox{\boldmath{$0$}}}

\newcommand{\bPhi}{\mbox{\boldmath{$\Phi$}}}

\newcommand{\bvarphi}{\mbox{\boldmath{$\varphi$}}}

\newtheorem{theorem}{Theorem}[section]
\newtheorem{remark}{Remark}

\newtheorem{lemma}[theorem]{Lemma}

\graphicspath{ {figures/} }

\begin{document}


\title{Design of MIMO Radar Waveforms based on $\ell_p$-Norm Criteria}

\author{
Ehsan~Raei,~\IEEEmembership{Student Member,~IEEE,}
Mohammad~Alaee-Kerahroodi,~\IEEEmembership{Member,~IEEE,}\\
Prabhu~Babu,
and~M.R.~Bhavani~Shankar,~\IEEEmembership{Senior~Member,~IEEE}
}


\maketitle

\begin{abstract}

Multiple-input multiple-output (MIMO) radars transmit a set of sequences that exhibit small cross-correlation sidelobes, to enhance sensing performance by separating them at the matched filter outputs. The waveforms also require small auto-correlation sidelobes to avoid masking of weak targets by the range sidelobes of strong targets and to mitigate deleterious effects of distributed clutter. In light of these requirements, in this paper, we design a set of phase-only (constant modulus) sequences that exhibit near-optimal properties in terms of Peak Sidelobe Level (PSL) and Integrated Sidelobe Level (ISL). At the design stage, we adopt weighted $\ell_p$-norm of  auto- and cross-correlation sidelobes as the objective function and minimize it for a general $p$ value, using block successive upper bound minimization (BSUM). Considering the limitation of radar amplifiers, we design unimodular sequences which make the design problem non-convex and NP-hard. To tackle the problem, in every iteration of the BSUM algorithm,
we introduce different local approximation functions 
and optimize them concerning a block, containing a code entry or a code vector. 
The numerical results show that the performance of the optimized set of sequences outperforms the state-of-the-art counterparts, in both terms of PSL values and computational time. 
\end{abstract}


\IEEEpeerreviewmaketitle

\section{Introduction}

A complex problem in radar pulse compression (intra-pulse modulation) is the design of waveforms exhibiting small \gls{PSL}. 
\gls{PSL} shows the maximum auto-correlation sidelobe of a transmit waveform in a typical \gls{SISO}/\gls{SIMO}, or phased-array radar system. If this value is not small, then either a false detection or a miss detection may happen, based on the way the \gls{CFAR} detector is tuned \cite{7771665}. 
In \gls{MIMO} radars, \gls{PSL} minimization is more complex since the cross-correlation sidelobes of transmitting set of sequences need to be also considered. Small value in cross-correlation sidelobes helps the radar receiver to separate the transmitting waveforms and form a \gls{MIMO} virtual array.

Similar properties hold for \gls{ISL} of transmitting waveforms where in case of \gls{SISO}/\gls{SIMO} or phased-array radars, the energy of auto-correlation sidelobes should be small to mitigate the deleterious effects of distributed clutter.
In solid state-based weather radars, \gls{ISL} needs to be small to enhance reflectively estimation and improve the performance of hydrometer  classifier \cite{8952671}. In \gls{MIMO} radar systems, \gls{ISL} shows the energy leakage of different waveforms in addition to the energy of non-zero auto-correlation sidelobes. 
Indeed, correlation sidelobes are a form of self-noise that  
reduce the effectiveness of transmitting waveforms in every radar system \cite{6012522}.  
In a \gls{MIMO} radar system, different multiplexing schemes can be used to create zero values for cross-correlations of the transmitting waveforms, \gls{FDM}, \gls{DDM}, and \gls{TDM} as some examples \cite{7060251}. 
Currently, \gls{TDM}-\gls{MIMO} radars are commercialized in the automotive industry with a variety of functionalities from de-chirping and Doppler processing to angle estimation and tracking \cite{SWRA554_V2, 9272873}.
However, \gls{CDM}-\gls{MIMO} is the next step of the industry, which  can use more efficiently the available resources (time and frequency)  \cite{imec_V2}. 


In this paper, we devise a method called 
\gls{WeBEST} to design transmitting waveforms for \gls{CDM}-\gls{MIMO} radars. To this end, we adopt the weighted $\ell_p$-norm of auto- and cross-correlation sidelobes as the objective function and minimize it under  \gls{CP} and \gls{DP} constraints.
The weighting and $p$ values in the provided formulation create a possibility for   intelligent transmission   based  on  prevailing environmental conditions, where can select appropriate $p$ based on presence of distributed clutter or strong target \cite{6875736,7746567,8961364,9266513}.
For example, choosing $p \to 0$ and minimizing the  $\ell_p$-norm of auto- and cross-correlation sidelobes, a set of sequences with sparse sidelobes will be obtained. With $p=2$,   the resulting optimized set of sequences will have small \gls{ISL} value which performs  well in the presence of clutter. Further, by minimizing the 
 $\ell_p$-norm when $p \rightarrow  +\infty$,  the optimized set of sequence will have small \gls{PSL} and are well suited for enhancing the detection of point targets. 

\subsection{Background and Related Works}
{\sl Waveform design based on sidelobe reduction in \gls{SISO}/\gls{SIMO} or phased-array radar systems:} 
Research into design of waveforms with small \gls{ISL} and \gls{PSL} values has significantly increased over the past decade for single waveform transmitting radar systems \cite{4749273,4838816,7093191,7362231,7605511,7529179,7967829,8168352,8926388}. In case of  \gls{ISL} minimization, several  optimization frameworks are proposed, including power method-like iterations, \gls{MM}, \gls{CD}, \gls{GD} and \gls{ADMM} to name a few 
\cite{4749273,4838816, 7093191,7362231,7529179,7605511,7967829,8168352,8926388}. 
Further, joint \gls{ISL} and \gls{PSL} minimization based on \gls{CD}  under \gls{DP} and \gls{CP} constraints is proposed in \cite{7967829}. In this paper $\ell_p$-norm  of auto-correlation sidelobes when $p \rightarrow  +\infty$ is considered for the initialization. 
Similarly, several papers have considered $\ell_p$-norm minimization to design waveform with small \gls{PSL} values. 
In \cite{7362231,7605511}, \gls{MM} based approach are proposed for $\ell_p$-norm minimization when $p\geq2$. 
Also, the authors in \cite{8168352} 
proposed a \gls{GD} based approach for $\ell_p$-norm minimization when $p$ is an even number, i.e., $p=2n, n \in \mathbb{Z}^+$. 
The results in \cite{7967829} depict that a methodology based on the   $\ell_p$-norm of auto-correlation sidelobes  by  gradually increasing $p$, provides sequences with smaller \gls{PSL} values comparing with the direct minimization of the \gls{PSL}. Motivated by this observation, this paper investigates $\ell_p$-norm minimization of auto- and cross-correlation functions  to obtain set of sequences with very small \gls{PSL} values for \gls{MIMO} radar systems. 

{\sl Waveform design based on sidelobe reduction in \gls{MIMO} radar systems:} 
In order to design set of sequences with small auto- and cross-correlation sidelobes, several approaches including  Multi-\gls{CAN}/Multi-Pe\gls{CAN} \cite{5072243}, Iterative Direct Search \cite{cui2017constant}, \gls{ISL}New \cite{8239862}, \gls{MM}-Corr \cite{7420715} and \gls{CD} \cite{8768085, raei2021spatial}, are proposed all considering the \gls{ISL} as the design metric. On the other hand, few papers have focused on \gls{PSL} minimization for \gls{MIMO} radars \cite{8706639,9290368}. 
In \cite{8706639}  a \gls{CD} based approach is proposed to directly minimize a weighted sum of \gls{PSL} and \gls{ISL} for \gls{MIMO} radars under \gls{DP} constraint. In \cite{9290368} a \gls{MM} based  approach is proposed to directly minimize the \gls{PSL} and design set of sequences for \gls{MIMO} radar systems. 
In the current study, we design set of sequences with very small \gls{PSL} values by minimizing  $\ell_p$-norm of auto- and cross-correlation sidelobes for a set of sequences which was not addressed previously in the literature. In contrast to the previous studies, we solve the problem for a general $p$ value ($p > 0$)
under \gls{DP} constraint, and solve it 
for $p\in(0,1] \cup p \geq 2$ under \gls{CP} constraint. Interestingly, the obtained \gls{PSL} values are close to the welch lower bound and fill the gap between the best of literature and the lower bound. 
\tablename{~\ref{tab:Clarify With Other}} compares the contributions of the proposed \gls{WeBEST} method with the state-of-the art approaches.

\begin{table}
	\centering
	\caption{The difference of the proposed method with the state of the art.}
	\begin{tabular}{c|c|c|c|c|c}	
		\hline
		\hline
		Paper & PSL & ISL  & $\ell_p$-norm & type & weight \\
		\hline
		\cite{7967829} & {\color{green} \checkmark} & {\color{green} \checkmark} & $p \geq 2$ & \gls{SISO} & {\color{red} $\times$}\\
		\hline
		\cite{7362231, 7605511} & {\color{green} \checkmark} & {\color{green} \checkmark} & $p \geq 2$ & \gls{SISO} & {\color{green} \checkmark}  \\
		\hline
		\cite{4749273} & {\color{red} $\times$} & {\color{green} \checkmark} & {\color{red} $\times$} & \gls{SISO} & {\color{green} \checkmark} \\
		\hline
		\cite{4838816, 8926388, 7529179} & {\color{red} $\times$} & {\color{green} \checkmark} & {\color{red} $\times$} &  \gls{SISO} & {\color{red} $\times$} \\
		\hline
		\cite{8168352} & {\color{green} \checkmark} & {\color{green} \checkmark} & $p \geq 2$ for $p$ even & \gls{SISO} &  {\color{green} \checkmark} \\
		\hline
		\cite{5072243, cui2017constant, 8239862, 7420715, 8768085, raei2021spatial} & {\color{red} $\times$} & {\color{green} \checkmark} & {\color{red} $\times$} & \gls{MIMO} & {\color{green} \checkmark}\\
		\hline
		\cite{8706639} & {\color{green} \checkmark} & {\color{green} \checkmark} & {\color{red} $\times$} & \gls{MIMO} & {\color{red} $\times$} \\
		\hline
		\cite{9290368} & {\color{green} \checkmark} & {\color{red} $\times$} & {\color{red} $\times$} & \gls{MIMO} & {\color{red} $\times$} \\
		\hline
		\gls{WeBEST} (\gls{DP}) & {\color{green} \checkmark} & {\color{green} \checkmark} & $p>0$ & \gls{MIMO} & {\color{green} \checkmark} \\
		\gls{WeBEST} (\gls{CP}) & {\color{green} \checkmark} & {\color{green} \checkmark} & $p\in(0,1] \cup p \geq 2$ & \gls{MIMO} & {\color{green} \checkmark} \\
		\hline
		\hline
	\end{tabular}
	\label{tab:Clarify With Other}
\end{table}

\subsection{Contributions}
The main contributions of the current article are summarized below.
\begin{itemize}
\item  {\sl Unified optimization framework:} 
We propose a unified framework based on \gls{BSUM} paradigm to solve a general $\ell_p$-norm of auto- and cross-correlation minimization problem under practical waveform design constraints which make the problem non-convex, non-smooth and NP-hard. While \gls{BSUM} offers a generic framework, the contribution of the paper lies in devising different solutions based on implementation complexity
and performance under a unified framework that solves the problem.
The proposed problem formulation includes $\ell_1$/$\ell_0$-norm of the auto-correlation sidelobe which relatively have lower number of local minima comparing with $l_2$-norm. Also, the local minima of those cost function would correspond to sequences with good auto-correlation sidelobe levels. For instance, in the simulation analysis we show that any local minima of $\ell_0$-norm  of auto-correlation would have many zeros (sparse auto-correlation) which can enhance the detection performance in the presence of distributed clutter.


\item  {\sl Entry- and vector-based solutions:} In each iteration of \gls{BSUM}, we propose two approaches, i.e, entry- and vector-based solutions. In the entry-based optimization, we formulate the problem with respect to a single variable; this enable us to find the critical points  and obtain the global optimum solution in each step.
For vector-based optimization we propose a solution based on \gls{GD}. This approach is faster than the entry-based method. However, the entry-based method has a better performance in terms of minimizing the objective function due to obtaining the global optimum solution in each step.  

\item  {\sl Trade-off and flexibility:}
By conducting thorough performance assessment, we propose a flexible tool to design set of sequences with different properties.
We show that the $\ell_p$-norm optimization framework provides the flexibly  of controlling optimization objective by choosing $p$, where $ p \to \infty$ leads to design set of waveforms with good \gls{PSL} property. Choosing $p \to 0$ leads to sparse auto- and cross-correlation and choosing $p = 2$ leads to design set of waveforms with good \gls{ISL} property.



\end{itemize}
 
We finally propose a direct solution for the discrete phase constraint using \gls{FFT}-based technique.

\subsection{Organization and Notations}
The rest of this paper is organized as follows. In Section \ref{sec:Problem Formulation and Optimization Framework}, we formulate the $\ell_p$-norm minimization for \gls{MIMO} radar systems, then we introduce the \gls{BSUM} method as the Optimization framework and finally we define the local approximation functions suitable for $\ell_p$-norm problem. We develop the \gls{BSUM} framework to solve the problem in Section \ref{sec:Proposed Method} and provide numerical experiments to verify the effectiveness of proposed algorithm in Section \ref{sec:Numerical Results}.
\paragraph*{Notations} This paper uses lower-case and upper-case boldface for vectors ($\ba$) and matrices ($\bA$) respectively. The set of complex and positives integer numbers are denoted by $\mathbb{C}$ and $\mathbb{Z}^+$  respectively. The transpose, conjugate transpose and sequence reversal are denoted by the $(.)^T$, $(.)^H$ and $(.)^r$ symbols respectively. Besides the Frobenius norm, $\ell_p$ norm, absolute value and round operator are denoted by $\norm{.}_F$, $\norm{.}_p$, $|.|$ and $\lfloor . \rceil$, respectively. For any complex number $a$, $\Re(a)$ and $\Im(a)$ denotes the real and imaginary part respectively. The letter $j$ represents the imaginary unit (i.e., $j=\sqrt{-1}$), while the letter $(i)$ is use as step of a procedure. Finally $\odot$ and $\circledast$ denotes the Hadamard product and cross-correlation operator respectively.

\section{Problem Formulation and Optimization Framework}\label{sec:Problem Formulation and Optimization Framework}
We consider a narrow-band \gls{MIMO} radar system with $M$ transmitters and each transmitting a sequence of length $N$ in the fast-time domain. Let the matrix $\bX \in \mathbb{C}^{M \times N}$ denote the set of transmitted sequences in baseband,
whose the $m^{th}$ row indicates the $N$ samples of $m^{th}$ transmitter while the $n^{th}$ column indicates the $n^{th}$ time-sample across the $M$ transmitters. Let $\mathbf{x}_m \triangleq [x_{m,1}, x_{m,2}, \dots, x_{m,N}]^T \in \mathbb{C}^{N}$ be the transmitted signal from $m^{th}$ transmitter. The aperiodic cross-correlation of $\mathbf{x}_m$ and $\mathbf{x}_l$ is defined as,
\begin{equation}\label{eq:cross_correlation}
	r_{m,l}(k) \triangleq (\mathbf{x}_m \circledast \mathbf{x}_l)_k = \textstyle \sum_{n=1}^{N-k} x_{m,n}x_{l,n+k}^*,
\end{equation}
where $m,l \in \{1,\dots,M_t\}$ are the transmit antennas indices and $k \in \{-N+1,\dots,N-1\}$ is the lag of cross-correlation. If $m = l$, \eqref{eq:cross_correlation} represents the aperiodic auto-correlation of signal $\mathbf{x}_m$. The zero lag of auto-correlation ($r_{m,m}(0)$) represent the mainlobe of the matched filter output. Also $|r_{m,m}(0)|$ contains the energy of sequence which for constant modulus sequences is equal to $N$. The other lags ($k \neq 0$) are referred to the sidelobes. The weighted $\ell_p$-norm of auto- and cross correlation in \gls{MIMO} radar can be written as, 
\begin{equation}\label{eq:wLp_Norm}
	\left(\textstyle \sum_{m=1}^{M}\sum_{l=1}^{M}\sum_{k=-N+1}^{N-1}|w_k r_{m,l}(k)|^p -  M(w_0N)^p\right)^\frac{1}{p},
\end{equation}
where, $0 \leq w_k \leq 1$.
The $M(w_0N)^p$ term in \eqref{eq:wLp_Norm} is the weighted $\ell_p$-norm of the mainlobes, where $\sum_{m=1}^{M}|w_0 r_{m,m}(0)|^p = M(w_0N)^p$.
Since the term $M(w_0N)^p$ in \eqref{eq:wLp_Norm} is constant, the weighted $\ell_p$-norm minimization can be equivalently written as,
\begin{equation}\label{eq:P}
	\mathcal{P}
	\begin{dcases}
	\min_{\bX} 	& f(\bX) \triangleq \textstyle \sum_{m=1}^{M}\sum_{l=1}^{M}\sum_{k=-N+1}^{N-1}|w_k r_{m,l}(k)|^p\\
	s.t 	    & x_{m,n} \in {\mathcal{X}}_{\infty} \quad \text{or} \quad {\mathcal{X}}_L,\\
\end{dcases}
\end{equation}
where, ${\mathcal{X}}_{\infty}$ and ${\mathcal{X}}_L$ indicating the unimodular and discrete phase with $L$ alphabet size sequences. More precisely, we consider $\mathcal{X}_{\infty} = \{e^{j\phi}|\phi \in \Omega_{\infty}\}$ and $\mathcal{X}_L = \{e^{j\phi}|\phi \in \Omega_L\}$, where $\Omega_{\infty} \triangleq (-\pi, \pi]$ and $\Omega_L \triangleq \{0, \frac{2\pi}{L}\, \dots, \frac{2\pi(L-1)}{L}\}$. 
The unimodular and discrete phase are equality constraint and they are not an affine set. Therefore the optimization problem not only is non-convex, but also multi-variable and NP-hard in general. Besides, due to the parameter $p$, in general dealing directly with $f(\bX)$ is complicated. 
In the following we introduce \gls{BSUM} method to solve the optimization problem effectively.


\subsection{\gls{BSUM} framework}
The \gls{BSUM} algorithm 
includes algorithms that successively optimize particular upper-bounds or local approximation functions of the original objectives in a block by block manner \cite{614066,7366709,1412048,9093027}. 
Let 
$\bX \triangleq [\mathbf{x}_1^T;\dots; \mathbf{x}_M^T] \in \mathbb{C}^{M \times N}$, where $\mathbf{x}_m, m = 1,\dots, M$ is the transmitted signal from $m^{th}$ transmitter. The following optimization problem,
\begin{equation}\label{eq:fopt}
\begin{dcases} 
\min_{\mathbf{x}} & f (\mathbf{x}_1, \mathbf{x}_2, \ldots, \mathbf{x}_M), \\
\text{s.t.} & \mathbf{x}_m \in {\mathcal{X}}_m, ~ m = 1, \ldots, M.
\end{dcases}
\end{equation}
can be iteratively solved  using the \gls{BSUM} technique, by finding the solutions of the following sub-problems for
$i = 0, 1, 2, \ldots $,
\begin{equation*}
    \begin{aligned}
    \mathbf{x}_1^{(i+1)} = &  \arg\displaystyle{\min_{\mathbf{x}_1 \in \mathcal{X}_1}} ~~ u_1(\mathbf{x}_1, \mathbf{x}_2^{(i)},\mathbf{x}_3^{(i)}, \ldots, \mathbf{x}_M^{(i)}),\\
    \mathbf{x}_2^{(i+1)} = & \arg\displaystyle{\min_{\mathbf{x}_2 \in \mathcal{X}_2}} ~~ u_2(\mathbf{x}_1^{(i+1)}, \mathbf{x}_2, \mathbf{x}_3^{(i)},\ldots, \mathbf{x}_M^{(i)}),\\
    \vdots\\
    \mathbf{x}_N^{(i+1)} = & \arg\displaystyle{\min_{\mathbf{x}_M \in \mathcal{X}_M}} ~~ u_n(\mathbf{x}_1^{(i+1)}, \mathbf{x}_2^{(i+1)}, \mathbf{x}_3^{(i+1)},\ldots, \mathbf{x}_M),
\end{aligned}
\end{equation*}
where $u_n$ is {\it local approximation} of the objective function. The \gls{BSUM} procedure consists of three steps as follows, 
\begin{itemize}
    \item We select a block.
    \item We find a local approximation function that locally approximates the objective function.
    \item At every iteration $i$, a single block, say $m = (i~ \text{mod} ~M) + 1$, is optimized by minimizing a approximation function of the selected block.
\end{itemize}

If at some point, the objective is not decreasing at every coordinate direction, then we have obtained the optimum $\bX^{\star} \equiv\bX^{(i+1)} \triangleq [{\mathbf{x}_1^{(i+1)}}^T, {\mathbf{x}_2^{(i+1)}}^T, \ldots, {\mathbf{x}_M^{(i+1)}}^T]$. The above framework is rather general, and leaves us the freedom of how to choose the index $m$ at $i$-th iteration.


\subsection{Choice of local approximation Functions}\label{sec:local approximation Functions}
The local approximation functions play an important role to simplify and efficiently solve the optimization problem. In the following, we introduce some local approximation functions which reduce the weighted $\ell_p$-norm problem of \eqref{eq:P} to simpler quadratic forms for $0 < p \leq 1$ and $p \geq 2$.

\subsubsection{local approximation Function for $p\geq2$}
In this case, one choice for local approximation function is using majorization function \cite{7366709}.
Let $u(\mathbf{x})$ be a majorization (minorization) function of $f(\mathbf{x})$ and $\mathbf{x}^{(i)}$ be the variable at $i^{(th)}$ iteration. This function must satisfy the following conditions \cite{7736116}, 
\begin{subequations}\label{eq:MM_Conditions}
\begin{align}
    &u(\mathbf{x}^{(i)}) = f(\mathbf{x}^{(i)}); \ \forall \mathbf{x}^{(i)} \in {\mathcal{X}}\\
    &u(\mathbf{x}) \geq f(\mathbf{x}) \ (\text{minorize:} \ u(\mathbf{x}) \leq f(\mathbf{x})); \ \forall \mathbf{x}, \in {\mathcal{X}}\\
    &\nabla u(\mathbf{x}^{(i)}) = \nabla f(\mathbf{x}^{(i)}); \ \forall \mathbf{x}^{(i)} \in {\mathcal{X}}\\
    &u(\mathbf{x}) \ \text{is continuous} \ \forall \mathbf{x}, \in {\mathcal{X}}.
\end{align}
\end{subequations}
When $p \geq 2$, $|w_kr_{m,l}(k)|^p$ can be majorized by the following function \cite{7362231},
\begin{equation}\label{eq:local approximation Function 2}
\eta_{mlk}|w_kr_{m,l}(k)|^2 + \psi_{mlk}|w_kr_{m,l}(k)| + \nu_{mlk}
\end{equation}
where,
\begin{equation}
\begin{aligned}
    \eta_{mlk} &\triangleq \frac{\tau^p + (p-1)|w_kr_{m,l}^{(i)}(k)|^p - p\tau|w_kr_{m,l}^{(i)}(k)|^{(p-1)}}{(\tau - |w_kr_{m,l}^{(i)}(k)|)^2}\\
    \psi_{mlk} &\triangleq p|w_kr_{m,l}^{(i)}(k)|^{(p-1)} - 2\eta_{mlk}|w_kr_{m,l}^{(i)}(k)|\\
    \nu_{mlk} &\triangleq \eta_{mlk}|w_kr_{m,l}^{(i)}(k)|^2 - (p-1)|w_kr_{m,l}^{(i)}(k)|^p
\end{aligned}
\end{equation}
and
\begin{equation}
    \tau \triangleq \textstyle \left(\sum_{-N-1}^{N-1}|w_kr_{m,l}^{(i)}(k)|^p\right)^{\frac{1}{p}}
\end{equation}
Furthermore, \eqref{eq:local approximation Function 2} can be majorized by \cite{7362231},
\begin{equation}\label{eq:local approximation Function 3}
\begin{aligned}
u(w_kr_{m,l}(k)) &\triangleq \eta_{mlk}|w_kr_{m,l}(k)|^2 \\
                 &+ \psi_{mlk}\Re \left\{w_k^* r_{m,l}^*(k) \frac{w_kr_{m,l}^{(i)}(k)}{|w_kr_{m,l}^{(i)}(k)|}\right\} + \nu_{mlk}
\end{aligned}
\end{equation}
Thus, the quadratic local approximation function of $f(\bX)$ for $p \geq 2$ is,
\begin{equation}
    u(\bX) \triangleq \textstyle \sum_{m=1}^{M}\sum_{l=1}^{M}\sum_{k=-N+1}^{N-1} u(w_kr_{m,l}(k))
\end{equation}

\subsubsection{local approximation Function for $0<p\leq1$} 
$f(\bX)|_{p \to 0}$ denotes the number of non-zero elements of auto- and cross-correlation. In order to avoid the singularity problem to obtain the derivative of $f(\bX)$, we replace $f(\bX)|_{p \to 0}$ with {\it smooth approximation} functions $g_h(\bX) \triangleq \sum_{m=1}^{M}\sum_{l=1}^{M}\sum_{k=-N+1}^{N-1} g_h(w_kr_{m,l}(k))$, $h \in \{1,2,3\}$, where \cite{7017587},
\begin{equation}
    \begin{aligned}
	g_1(r_{m,l}(k)) &\triangleq |w_kr_{m,l}(k)|^p, \quad 0 < p \leqslant 1,\\
	g_2(r_{m,l}(k)) &\triangleq \frac{\ln(1+\frac{|w_kr_{m,l}(k)|^p}{p})}{\ln(1+\frac{1}{p})}, \quad p > 0,\\
	g_3(r_{m,l}(k)) &\triangleq 1-e^{-\frac{|w_kr_{m,l}(k)|^p}{p}}, \quad p > 0.
	\end{aligned}
\end{equation}
The aforementioned {smooth} approximation functions may simplify the optimization problem, but they still have an order $p$. This means that is not yet easy to optimize the above { smooth} functions. To find a local approximation function, notice that each of the above { smooth} approximations can be majorized with the following simpler quadratic function \cite{7017587},
\begin{equation}\label{eq:v_h}
	v_h(w_kr_{m,l}(k)) \triangleq \gamma_{hmlk}|w_kr_{m,l}(k)|^2 + \mu_{hmlk}
\end{equation}
where, the coefficients $\gamma_{hmlk}$ and $\mu_{hmlk}$ can be obtained by solving the following system of equation \cite{7017587},
\begin{equation}
\begin{aligned}
	g_h(w_kr_{m,l}^{(i)}(k)) = & v_h(w_kr_{m,l}^{(i)}(k)) \\
	\frac{\partial g_h(w_kr_{m,l}^{(i)}(k))}{\partial |w_kr_{m,l}^{(i)}(k)|} = & 2\gamma_{hmlk}|w_kr_{m,l}^{(i)}(k)|.
\end{aligned}
\end{equation}
resulting in,
\begin{equation}\label{eq:mu_hmlk}
\begin{aligned}
	&\mu_{hmlk} = g_h(w_kr_{m,l}^{(i)}(k)) - \gamma_h|w_kr_{m,l}^{(i)}(k)|^2 \\
	&\gamma_{hmlk} = \frac{\partial g_h(w_kr_{m,l}^{(i)}(k))}{\partial |w_kr_{m,l}^{(i)}(k)|} \times \frac{1}{2|w_kr_{m,l}^{(i)}(k)|},
\end{aligned}
\end{equation}
The quadratic functions in \eqref{eq:v_h}, \eqref{eq:mu_hmlk} are non-differentiable and singular when $w_kr_{m,l}(k) = 0$. 
A solution suggested in \cite{7017587} 
is to incorporate a small $\epsilon > 0$
that avoids this singularity issue and use the smooth approximation functions $g_h^{\epsilon}(w_kr_{m,l}(k))$ and $\gamma_{hmlk}^\epsilon$ which are written in \tablename{~\ref{tab:smooth_and_local approximation}}. In this table, $\mu_{hmlk}^\epsilon$ is not reported, since it is a constant term and does not affect the optimization procedure.

\begin{table*}
	\centering
\caption[]{The {smooth} approximation function of $l_p$-norm and correspond local approximation function when $0<p<1$}
	\begin{tabular}{c|c}	
		\hline
		\hline
		Smooth approximation functions ($g_h^\epsilon(w_kr_{m,l}(k))$) & Coefficients of majorization functions ($\gamma_{hmlk}^\epsilon$) \eqref{eq:v_h}\\
		\hline
        $\begin{dcases}
    	\frac{p}{2}\epsilon^{p-2}|w_kr_{m,l}(k)|^2	& |w_kr_{m,l}(k)| \leqslant \epsilon \\
    	|w_kr_{m,l}(k)|^p - (1-\frac{1}{p}) \epsilon^p	    & |w_kr_{m,l}(k)| > \epsilon\\
    	\end{dcases}$
        &
        $\begin{dcases}
    	\frac{p\epsilon^{(p-2)}}{2}	& |w_kr_{m,l}(k)| \leqslant \epsilon \\
    	\frac{p|w_kr_{m,l}(k)|^{(p-2)}}{2}	    & |w_kr_{m,l}(k)| > \epsilon\\
    	\end{dcases}$
        \\    
        $\begin{dcases}
    	\frac{|w_kr_{m,l}(k)|^2}{2\epsilon(p+\epsilon)\ln(\frac{p+1}{p})}	& |w_kr_{m,l}(k)| \leqslant \epsilon \\
    	\frac{\ln(\frac{p+|w_kr_{m,l}(k)|}{p}) - \ln(\frac{p+\epsilon}{p}) + \frac{\epsilon}{2(p+\epsilon)}}{\ln(\frac{p+1}{p})}	    & |w_kr_{m,l}(k)| > \epsilon\\
    	\end{dcases}$
    	&
    	$\begin{dcases}
    	\frac{0.5}{\epsilon(p+\epsilon)\ln(\frac{p+1}{p})}	& |w_kr_{m,l}(k)| \leqslant \epsilon \\
    	\frac{0.5}{\ln(\frac{p+1}{p})|w_kr_{m,l}(k)|(|w_kr_{m,l}(k)|+p)}	    & |w_kr_{m,l}(k)| > \epsilon\\
    	\end{dcases}$
    	\\
        $\begin{dcases}
    	\frac{e^{-\frac{\epsilon}{p}}}{2p\epsilon}|w_kr_{m,l}(k)|^2	& |w_kr_{m,l}(k)| \leqslant \epsilon \\
    	-e^{-\frac{|w_kr_{m,l}(k)|}{p}} + (1+\frac{\epsilon}{2p})e^{-\frac{\epsilon}{p}}	    & |w_kr_{m,l}(k)| > \epsilon\\
    	\end{dcases}$
    	&
    	$\begin{dcases}
    	\frac{e^{-\frac{\epsilon}{p}}}{2p\epsilon}	& |w_kr_{m,l}(k)| \leqslant \epsilon \\
    	\frac{e^{-\frac{|w_kr_{m,l}(k)|}{p}}}{2p|w_kr_{m,l}(k)|}	    & |w_kr_{m,l}(k)| > \epsilon\\
    	\end{dcases}$
	\\
		\hline
		\hline
	\end{tabular}
	\label{tab:smooth_and_local approximation}
\end{table*}

Thus the majorization function of $g_h^{\epsilon}(\bX)$ with $0 < p \leq 1$ is,
\begin{equation}
	v_h^{\epsilon}(\bX) \triangleq \textstyle \sum_{m=1}^{M}\sum_{l=1}^{M}\sum_{k=-N+1}^{N-1} v_h^{\epsilon}(w_kr_{m,l}(k))
\end{equation}

In the following, we propose a framework to solve the optimization problem based on \gls{BSUM} method and we consider cyclic rule to update the waveform. In this framework the block can be either one vector ($\mathbf{x}_m$) or one entry ($x_{m,n}$) of the waveform matrix $\bX$. In the following we propose two methods based on entry and vector  optimization.

\section{Proposed Method}\label{sec:Proposed Method}
\gls{BSUM} optimization methodology requires the problem in $\mathcal{P}$ be written in a simplified form with respect to one block while others are held fixed. In this regard, let $\mathbf{x}_t$ ($t \in \{1, \dots, M\}$) be the only variable block, while other blocks are held fixed and stored in the matrix $\bX_{-t} \triangleq [\mathbf{x}_1^T;\dots; \mathbf{x}_{t-1}^T; \bzero^T; \mathbf{x}_{t+1}^T; \dots; \mathbf{x}_M^T] \in \mathbb{C}^{M \times N}$, where $\bzero^T$ denotes $1\times N$ vector which all the entries are equal to zero. In this case, the objective functions $f(\bX)$, $g_h^{\epsilon}(\bX)$, $v_h^{\epsilon}(\bX)$ and $u(\bX)$ can be decomposed as independent term, auto- and cross-correlation terms of $\mathbf{x}_t$. for example $f(\bX)$ can be written as follows,
\begin{equation}
	f(\bX) = f_m(\bX_{-t}) + f_{au}(\mathbf{x}_t) + f_{cr}(\mathbf{x}_t, \bX_{-t})
\end{equation}
where, $f_m(\bX_{-t})$ denotes the independent term of $\mathbf{x}_t$, while $f_{au}(\mathbf{x}_t)$ and $f_{cr}(\mathbf{x}_t, \bX_{-t})$ denotes the $\mathbf{x}_t$ dependent terms of auto- and cross-correlation respectively. After some mathematical manipulations the functions $f(\bX)$, $g_h^{\epsilon}(\bX)$, $v_h^{\epsilon}(\bX)$ and $u(\bX)$ can be decomposed as reported in \tablename{~\ref{tab:w.r.t vector}}.

\begin{table*}
	\centering
\caption[]{Decomposition of functions $f(\bX)$, $g_h^{\epsilon}(\bX)$, $v_h^{\epsilon}(\bX)$ and $u(\bX)$.}
	\begin{tabular}{c|c|c|c}	
		\hline
		\hline
		Function 
		& 
		\shortstack{Independent term \\ ($f_m,g_{h,m}^{\epsilon},v_{h,m}^{\epsilon},u_m(\bX_{-t})$)}
		& 
		\shortstack{ Auto-correlation term \\ ($f_{au},g_{h,au}^{\epsilon},v_{h,au}^{\epsilon},u_{au}(\mathbf{x}_t)$)}
		& 
		\shortstack{ Cross-correlation term \\ ($f_{cr},g_{h,cr}^{\epsilon},v_{h,cr}^{\epsilon},u_{cr}(\mathbf{x}_t,\bX_{-t})$)}
		\\ \hline
        $f(\bX)$ 
        & 
        $\sum_{\substack{{m,l=1}\\{m,l \neq t}}}^{M}\sum_{k=-N+1}^{N-1}|w_k r_{m,l}(k)|^p$
        & 
        $\sum_{k=-N+1}^{N-1} |w_k r_{t,t}(k)|^p$
        & 
        $2\sum_{\substack{{l=1}\\{l \neq t}}}^{M}\sum_{k=-N+1}^{N-1}|w_k r_{t,l}(k)|^p$
        \\ \hline
        $g_h^{\epsilon}(\bX)$ 
        &
        $\sum_{\substack{{m,l=1}\\{m,l \neq t}}}^{M}\sum_{k=-N+1}^{N-1}g_h^{\epsilon}(w_kr_{m,l}(k))$
        & 
        $\sum_{k=-N+1}^{N-1} g_h^{\epsilon}(w_kr_{t,t}(k))$
        & 
        $2\sum_{\substack{{l=1}\\{l \neq t}}}^{M}\sum_{k=-N+1}^{N-1}g_h^{\epsilon}(w_kr_{t,l}(k))$
        \\ \hline
        $v_h^{\epsilon}(\bX)$ 
        &
        $\sum_{\substack{{m,l=1}\\{m,l \neq t}}}^{M}\sum_{k=-N+1}^{N-1} v_h^{\epsilon}(w_kr_{m,l}(k))$
        & 
        $\sum_{k=-N+1}^{N-1} v_h^{\epsilon}(w_kr_{t,t}(k))$
        & 
        $2\sum_{\substack{{l=1}\\{l \neq t}}}^{M}\sum_{k=-N+1}^{N-1} v_h^{\epsilon}(w_kr_{t,l}(k))$
        \\ \hline
        $u(\bX)$ 
        &
        $\sum_{\substack{{m,l=1}\\{m,l \neq t}}}^{M}\sum_{k=-N+1}^{N-1} u(w_kr_{m,l}(k))$
        & 
        $\sum_{k=-N+1}^{N-1} u(w_kr_{t,t}(k))$
        & 
        $2\sum_{\substack{{l=1}\\{l \neq t}}}^{M}\sum_{k=-N+1}^{N-1} u(w_kr_{t,l}(k))$ 
        \\
		\hline
		\hline
	\end{tabular}
	\label{tab:w.r.t vector}
\end{table*}

\begin{table*}
	\centering
\caption[]{Expressing the auto- and cross-correlation terms of $f(\bX)$, $v_h^{\epsilon}(\bX)$ and $u(\bX)$ with respect to $x_{t,d}$.}
	\begin{tabular}{c|c|c}	
		\hline
		\hline
		Function 
		& 
		\shortstack{ Auto-correlation term with respect to $x_{t,d}$\\ ($f_{au},g_{h,au}^{\epsilon},v_{h,au}^{\epsilon},u_{au}(\mathbf{x}_t)$)}
		& 
		\shortstack{ Cross-correlation term with respect to $x_{t,d}$\\ ($f_{cr},g_{h,cr}^{\epsilon},v_{h,cr}^{\epsilon},u_{cr}(\mathbf{x}_t,\bX_{-t})$)}
		\\ \hline
        $f(\bX)$ 
        & 
        $\sum_{k=-N+1}^{N-1} |c_{ttdk} + a_{ttdk}x_{t,d} + b_{ttdk}x_{t,d}^*|^p$
        & 
        $2\sum_{\substack{{l=1}\\{l \neq t}}}^{M}\sum_{k=-N+1}^{N-1}|c_{tldk} + a_{tldk}x_{t,d}|^p$
        \\ \hline
        $v_h^{\epsilon}(\bX)$ 
        & 
        $\sum_{k=-N+1}^{N-1} \gamma_{httk}|c_{ttdk} + a_{ttdk}x_{t,d} + b_{ttdk}x_{t,d}^*|^2 + \mu_{httk}$
        & 
        $2\sum_{\substack{{l=1}\\{l \neq t}}}^{M}\sum_{k=-N+1}^{N-1} \gamma_{htlk}|c_{tldk} + a_{tldk}x_{t,d}|^2 + \mu_{htlk}$
        \\ \hline
        $u(\bX)$ 
        & 
        \shortstack{ $\sum_{k=-N+1}^{N-1}( \eta_{httk}|c_{ttdk} + a_{ttdk}x_{t,d} + b_{ttdk}x_{t,d}^*|^2 + $\\ $\psi_{httk}\Re \left\{(c_{ttdk} + a_{ttdk}x_{t,d} + b_{ttdk}x_{t,d}^*)^* \frac{w_kr_{t,t}^{(i)}(k)}{|w_kr_{t,t}^{(i)}(k)|}\right\} + \nu_{httk})$}
        & 
        \shortstack{ $2\sum_{\substack{{l=1}\\{l \neq t}}}^{M}\sum_{k=-N+1}^{N-1}( \eta_{htlk}|c_{tldk} + a_{tldk}x_{t,d}|^2 + $\\ $\psi_{htlk}\Re \left\{ (c_{tldk} + a_{tldk}x_{t,d})^* \frac{w_kr_{t,l}^{(i)}(k)}{|w_kr_{t,l}^{(i)}(k)|}\right\} + \nu_{htlk})$}
        \\
		\hline
		\hline
	\end{tabular}
	\label{tab:w.r.t entry}
\end{table*}

\subsection{Entry optimization}
In this case, we consider each entry of $\bX$ as block of \gls{BSUM} framework. Then, we select an entry as the only variable while keeping the others fixed. Thus, to express the problem with respect to the selected variable $x_{t,d}$, we follow these two steps:
\begin{itemize}
    \item We pick the $t^{th}$ transmitter then express the problem with respect to that transmitter.
    \item We pick the $d^{th}$ sample of the selected transmitter then express the problem with respect to that sample.
\end{itemize}
Let $x_{t,d}$ ($t \in \{1, \dots, M\}$ and $d \in \{1, \dots, N\}$)  be the only entry variable of vector $\mathbf{x}_t$ while other entries are held fixed and stored in vector $\mathbf{x}_{t,-d} \triangleq [x_{t,1}, \dots, x_{t,d-1}, 0, x_{t,d+1}, \dots, x_{t,N}]^T \in \mathbb{C}^N$. Therefore, the auto- and cross- correlation terms of functions $f(\bX)$, $v_h^{\epsilon}(\bX)$ and $u(\bX)$ can be obtained based on the only variable as reported in \tablename{~\ref{tab:w.r.t entry}} (see Appendix \ref{app:1} for more details about obtaining those) \footnote{Since in optimization procedure we do not deal directly with $g_h^{\epsilon}(\bX)$, we do not express it with respect to $x_{t,d}$.}.

Herein, we substitute $x_{t,d}$ with $e^{j\phi}; \phi \in \Omega_{\infty}$ to consider the unimodularity constraint directly in the objective function. In this case, the problem boils down to the following optimization problem (see Appendix \ref{app:2}),
\begin{equation}\label{eq:Pe_v}
	\mathcal{P}_{e, (0 < p \leq 1)}
	\begin{dcases}
	\min_{\phi} 	& v_h^{\epsilon}(\phi)\\
	s.t 	        & \phi \in \Omega_{\infty}\\
\end{dcases},
	\mathcal{P}_{e, (p \geq 2)}
	\begin{dcases}
	\min_{\phi} 	& u(\phi)\\
	s.t 	        & \phi \in \Omega_{\infty},\\
\end{dcases}
\end{equation}
where,
\begin{equation} 
\begin{aligned}\label{eq:Coeff_ui_vi}
    v_{h}^{\epsilon}(\phi) &\triangleq \sum_{n=-2}^{2} v_{h,n} e^{jn\phi},~~
    u(\phi) &\triangleq \Re\left\{\sum_{n=-2}^{2} u_n e^{jn\phi}\right\},
\end{aligned}
\end{equation}
and the coefficients $v_{h,n}$ and $u_n$ are given in Appendix \ref{app:2}.

The solution for $\mathcal{P}_{e, (0 < p \leq 1)}$ and $\mathcal{P}_{e, (p \geq 2)}$ will be obtained by finding the critical points of the problem and subsequently selecting the one that minimizes the objective. As $v_h^{\epsilon}(\phi)$ and $u(\phi)$ are differentiable and periodic functions over interval $[-\pi,\pi)$, the critical points of  $\mathcal{P}_{e, (0 < p \leq 1)}$ and $\mathcal{P}_{e, (p \geq 2)}$ contain the solutions to $\frac{dv_h^{\epsilon}(\phi)}{d\phi} = 0$ and $\frac{du(\phi)}{d\phi} = 0$, for $\phi \in \Omega_{\infty}$. In this regards, the derivative of $v_{h}^{\epsilon}(\phi)$ and $u(\phi)$ can be obtained by,
\begin{equation} \label{eq:d_ui_vi}
    v_h^{\epsilon'}(\phi) = \sum_{n=-2}^{2} jn v_{h,n} e^{jn\phi},~~
    u'(\phi) = \Re\left\{\sum_{n=-2}^{2} jn u_n e^{jn\phi}\right\},
\end{equation}
Considering $\cos(\phi) = {(1-\tan^2(\frac{\phi}{2}))}/{(1+\tan^2(\frac{\phi}{2}))}$, $\sin(\phi) = {2\tan(\frac{\phi}{2})}/{(1+\tan^2(\frac{\phi}{2}))}$ and using the change of variable $z \triangleq \tan(\frac{\phi}{2})$, it can be shown that finding the roots of $\frac{dv_h^{\epsilon}(\phi)}{d\phi} = 0$ and $\frac{du(\phi)}{d\phi} = 0$ are equivalent to find the roots of the following $4$ degree real polynomials (see Appendix \ref{app:3} for details),
\begin{equation}\label{eq:dfu_root}
 	\textstyle \sum_{k=0}^{4} q_{h,k}z^k = 0, \quad \sum_{k=0}^{4} s_kz^k = 0,
\end{equation}
respectively, where the coefficients are given in Appendix \ref{app:3}. 
We only admit the real roots for \eqref{eq:dfu_root}. Let us assume that $z_{v,k}$ and $z_{u,k}$, $k=\{1,\dots,4\}$ are the roots of $\sum_{k=0}^{4} q_{h,k}z^k = 0$ and $\sum_{k=0}^{4} s_kz^k = 0$ respectively. Hence, the critical points of $\mathcal{P}_{e, (0 < p \leq 1)}$ and $\mathcal{P}_{e, (p \geq 2)}$ can be expressed as, 
\begin{equation}\label{eq:phi_critical_points}
\begin{aligned}
    \Omega_v &= \left\{2\arctan{(z_{v,k})} | \Im(z_{v,k})=0 \right\}, \\
    \Omega_u &= \left\{2\arctan{(z_{u,k})} | \Im(z_{u,k})=0 \right\}
\end{aligned}
\end{equation}
respectively. Therefore, the optimum solution for $\mathcal{P}_{e, (0 < p \leq 1)}$ and $\mathcal{P}_{e, (p \geq 2)}$ are,
\begin{equation}\label{eq:optimum_phi_e}
\begin{aligned}
	\phi_v^{\star} =& \arg\min_{\phi} \left \{v_h^{\epsilon}(\phi) | \phi \in \Omega_v \right \},\\
	\phi_u^{\star} =& \arg\min_{\phi} \left \{u(\phi) | \phi \in \Omega_u \right \}.
\end{aligned}
\end{equation}
respectively. Subsequently the optimum solution for $x_{t,d}$ are, $x_{t,d}^{(i)} = e^{j\phi_v^{\star}}$ and $x_{t,d}^{(i)} = e^{j\phi_u^{\star}}$ respectively.

\begin{remark}
Since, $v_h^{\epsilon}(\phi)$ and $u(\phi)$ are functions of $\cos{\phi}$ and $\sin{\phi}$, it is periodic, real and differentiable. Therefore, it has at least two extrema and hence its derivative has at least two real roots; thus $\Omega_v$ and $\Omega_u$ never become a null set. As a result in each iteration, the problem has a solution and never becomes infeasible.
\end{remark}

\subsubsection{Discrete phase optimization}
The $\ell_p$-norm minimization can be solved directly for $0<p<\infty$ under discrete phase constraint. 
In this case all the discrete points lie on the boundary of the optimization problem; hence, all of them are critical points for the problem. Therefore, one approach for solving this problem is exhaustive search. In this method all the possible values of the objective function $f(\phi)$ over the set $\Omega_L = \left \{\phi_0,\phi_1,\dots,\phi_{L-1}\right \} \in \left \{0, \frac{2\pi}{L}, \dots, \frac{2\pi(L-1)}{L} \right \}$ are obtained and the phase minimizing the objective function is chosen. This method is too expensive in terms of complexity. However, for \gls{MPSK} alphabet, an elegant solution can be obtained by using 
\gls{FFT}
as detailed below. 

It can be shown that the $\ell_p$-norm of auto- and cross-correlations can be written with respect to alphabet indices as (see Appendix \ref{app:4} for details),
\begin{equation}\label{eq:P_l}
\begin{aligned}
\mathcal{P}_d
	\begin{dcases}
    \arg \displaystyle{\min_{l'=1,\dots,L}} &\{f(l') = f(\bX_{-t}) + \\
    &\textstyle 2\sum_{\substack{{l=1}\\{l \neq t}}}^{M}\sum_{k=-N+1}^{N-1} | \mathcal{F}_L\{a_{tldk}, c_{tldk}\}|^p + \\
    &\textstyle \sum_{k=-N+1}^{N-1} |\mathcal{F}_L\{a_{ttdk}, c_{ttdk}, b_{ttdk}\}|^p\}
    \end{dcases}
\end{aligned}
\end{equation} 
where, $l'$ are the indices of set $\Omega_L$ and $\mathcal{F}_L$ is the $L$ point \gls{DFT} operator. Due to aliasing phenomena when $L = 2$, the third term of $f(l')$ should be changed to $\sum_{k=-N+1}^{N-1} |w_k \mathcal{F}_L\{a_{ttdk} + b_{ttdk}, c_{ttdk}\}|^p$. Therefore, the optimum solution of \eqref{eq:P_l} is,
\begin{equation}\label{eq:l_star}
    l'^{\star} = \arg\displaystyle{\min_{l=1,\dots,L}} \left\{f(l')\right\}.
\end{equation} 
Hence, $\phi_d^{\star} = \frac{2\pi(l'^{\star}-1)}{L}$ and the optimum entry is $s_{t,d}^{(i)} = e^{j\phi_d^{\star}}$.

The summary of the proposed method, called \gls{WeBEST}-entry based design optimization framework is given by \textbf{Algorithm \ref{alg:entry_optimization}}, where, $x_{t,d}^{\star} = e^{j\phi^{\star}}$ is the optimized solution of optimization problem \eqref{eq:Pe_v}
($\phi^{\star} \in \{\phi_v^{\star}, \phi_u^{\star}, \phi_d^{\star}\}$). To obtain this solution, \gls{WeBEST}-e (entry optimization) considers a feasible set of sequences as the initial waveforms. Then, at each iteration, it selects $x_{t,d}^{(i)}$ as the variable and updates that with optimized ${x_{t,d}^{(i+1)}}$, denoted by ${x_{t,d}^{\star}}$. This procedure is repeated for other entries and is undertaken until all the entries are optimized at least once. After optimizing the $M N^{th}$ entry, the algorithm examines the convergence metric for the objective function. If the stopping criteria is not met the algorithm repeats the aforementioned steps. 
\begin{algorithm}[t]
	\caption{:\gls{WeBEST}-entry optimization framework}
	\label{alg:entry_optimization}
	\textbf{Input:} Initial set of feasible sequences, $\bX^{(0)}$.\\
	\textbf{Initialization:} $i := 0$.\\ 
	\textbf{Optimization:} 
	\begin{enumerate}
		\item {\bf while}, the stopping criteria is not met,  {\bf do}
		\item \hspace{5mm} $i := i+1$;
		\item \hspace{5mm} {\bf for} $t=1,\dots,M_t$ {\bf do}
		\item \hspace{10mm} {\bf for} $d=1,\dots,N$ {\bf do}
		\item \hspace{15mm} Optimize  $x_{t,d}^{(i)}$  and obtain $x^{\star}_{t,d}$;  
		\item \hspace{15mm} Update $x^{(i+1)}_{t,d} = x^{\star}_{t,d}$;
		\item \hspace{15mm} $\bX^{(i+1)} = \bX^{(i+1)}_{-(t,d)} |_{x_{t,d}=x^{(i+1)}_{t,d}}$;
		\item \hspace{10mm} {\bf end for}
		\item \hspace{5mm} {\bf end for}
		\item {\bf end while}
		\end{enumerate}
	\textbf{Output:} $\bX^{\star} = \bX^{(i+1)}$.
\end{algorithm} 

\subsection{Vector optimization 
} 
In this part, we propose the \gls{WeBEST}-vector optimization framework (\gls{WeBEST}-v) under continuous phase constraint. In this method, since we update 
a vector in every step, the convergence time is much faster than the entry optimization approach. In this regards, in the following, we propose an \gls{BSUM} based method where in each iteration an \gls{GD} method is deployed to update each transmitter waveform. 

Let $\bPhi \in \mathbb{R}^{M \times N}$ and $\bvarphi_t \in \mathbb{R}^N$ be the phases corresponding to the matrix $\bX$ ($\bPhi \triangleq \angle{\bX}$) and the vector variable $\mathbf{x}_t$ ($\bvarphi_t \triangleq \angle{\mathbf{x}_t}$) respectively. In general, the procedure starts with an initial solution ($\bPhi^{(0)}$), then at $i^{th}$ iteration, each block ($\bvarphi_t$) is updated by the following equation \cite{boyd2004convex}, 
\begin{equation}
    \bvarphi_t^{(i+1)} = \bvarphi_t^{(i)} + \delta^{(i)}\Delta\bvarphi_t^{(i)}
\end{equation}
where, $\delta^{(i)}$ and $\Delta\bvarphi_t^{(i)}$ are the \textit{step size} (\textit{step length}) and the \textit{search direction} at $i^{th}$ iteration, respectively. After updating all of the blocks, the phase matrix is updated by $\bPhi^{(i+1)} \triangleq [\bvarphi_1^{(i+1)},\dots,\bvarphi_M^{(i+1)}]^T$. In gradient descent method, the search direction is equal to the negative of the gradient i.e. $\Delta\bvarphi_t^{(i)} = -\nabla f(\bvarphi_t^{(i)})$, and a possible solution for step size is using \textit{backtracking line search} \cite{boyd2004convex}.




\textbf{Algorithm \ref{alg:vector_optimization}}, called 
\gls{WeBEST}-v shows the procedure of vector optimization of $\ell_p$-norm minimization. In this algorithm, matrix $\nabla \bPhi^{(i)} \in \mathbb{R}^{M \times N}$ contains the gradient of objective function with respect to sequence phases at $i^{th}$ iteration, i.e., $\nabla \bPhi^{(i)} \triangleq [\nabla_{\bvarphi_1}f(\bvarphi_1^{(i)}), \ldots, \nabla_{\bvarphi_M}f(\bvarphi_M^{(i)})]^T$. This procedure will be continued until the algorithm meet the stopping criteria\footnote{Please note that the \gls{WeBEST}-v is proposed for $2 \leq p < \infty$. For $0 < p \leq 1$, we can simply replace $f(\bvarphi_M^{(i)})$ with $g_h^{\epsilon}(\bvarphi_M^{(i)})$.}.
The algorithm requires calculation of 
the gradients of $\nabla_{\bvarphi_t} f(\bvarphi_t^{(i)})$ and $\nabla_{\bvarphi_t} g_h^{\epsilon}(\bvarphi_t^{(i)})$, which can be obtained using the following lemma.

\begin{algorithm}[t]
	\caption{: \gls{WeBEST}-vector optimization framework}
	\label{alg:vector_optimization}
	\textbf{Input:} $\bX^{(0)}$\\
	\textbf{Initialization:} $i := 0$, $\bPhi^{(i)} = \angle{\bX^{(0)}}$. 
	\begin{enumerate}
		\item {\bf while}, the stopping criteria is not met,  {\bf do}
		\item \hspace{5mm} {\bf for} $t := 1:M$ 
		\item \hspace{10mm} $\Delta\bvarphi_t^{(i)} := -\nabla_{\bvarphi_t}f(\bvarphi_t^{(i)})$;
		\item \hspace{10mm} obtain $\delta^{(i)}$ using backtracking line search;
		\item \hspace{10mm} $\bvarphi_t^{(i+1)} := \bvarphi_t^{(i)} + \delta^{(i)}\Delta\bvarphi_t^{(i)}$;
		\item \hspace{5mm} {\bf end for} 
		\item \hspace{10mm} $i:=i+1$;
		\item {\bf end while}
		\end{enumerate}
\end{algorithm}

\begin{lemma}
The gradient of $\nabla_{\bvarphi_t} f(\bvarphi_t^{(i)})$ and $\nabla_{\bvarphi_t} g_h^{\epsilon}(\bvarphi_t^{(i)})$ are equal to,
\par\noindent\small
\begin{equation}\label{eq:gradient_g}
\begin{aligned}
&\nabla_{\bvarphi_t} g_h^{\epsilon}(\bvarphi_t^{(i)}) = 4\Im\{\mathbf{x}_t^* \odot ((\bm{\varrho}_{htt}^{\epsilon^2} \odot (\mathbf{x}_t \circledast \mathbf{x}_t)) \circledast \mathbf{x}_t )_{k+N-1}\}\\
&+\textstyle 4\sum_{\substack{{l=1}\\{l \neq t}}}^{M}\Im\{\mathbf{x}_t^* \odot ((\bm{\varrho}_{htt}^{\epsilon^2} \odot (\mathbf{x}_l \circledast \mathbf{x}_t)^r) \circledast \mathbf{x}_l^* )_{k+N-1}\},
\end{aligned}
\end{equation}
\normalsize
\par\noindent\small
\begin{equation}\label{eq:gradient_f}
\begin{aligned}
 &\nabla_{\bvarphi_t}f(\bvarphi_t^{(i)}) = 4\Im\{\mathbf{x}_t^* \odot ((\bm{\vartheta}_{tt}^2 \odot (\mathbf{x}_t \circledast \mathbf{x}_t)) \circledast \mathbf{x}_t )_{k+N-1}\}\\
&+\textstyle 4\sum_{\substack{{l=1}\\{l \neq t}}}^{M}\Im\{\mathbf{x}_t^* \odot ((\bm{\vartheta}_{tl}^2 \odot (\mathbf{x}_l \circledast \mathbf{x}_t)^r) \circledast \mathbf{x}_l^* )_{k+N-1}\},
\end{aligned}
\end{equation}
\normalsize
where, $\bm{\varrho}_{htt}^{\epsilon} \triangleq [\varrho_{htt}^{\epsilon}(-N+1),\dots,\varrho_{htt}^{\epsilon}(N-1)]^T | \varrho_{htt}^{\epsilon}(k) \triangleq w_k\sqrt{\gamma_{httk}}$, $\bm{\varrho}_{htl}^{\epsilon} \triangleq [\varrho_{htl}^{\epsilon}(-N+1),\dots,\varrho_{htl}^{\epsilon}(N-1)]^T | \varrho_{htl}^{\epsilon}(k) \triangleq w_k\sqrt{\gamma_{htlk}}$, $\bm{\vartheta}_{tt} \triangleq [\vartheta_{tt}(-N+1),\dots,\vartheta_{tt}(N-1)]^T | \vartheta_{tt}(k) \triangleq w_k\sqrt{\nu_{ttk}}$ and $\bm{\vartheta}_{tl} \triangleq [\vartheta_{tl}(-N+1),\dots,\vartheta_{tl}(N-1)]^T | \vartheta_{tl}(k) \triangleq w_k\sqrt{\nu_{tlk}}$.
\end{lemma}

\textit{proof:}
Since the gradient of majorization/minorization function at point $\bvarphi_t^{(i)}$ is equal to the objective function, we can obtain the gradient of $\nabla_{\bvarphi_t} f(\bvarphi_t^{(i)})$ and $\nabla_{\bvarphi_t} g_h^{\epsilon}(\bvarphi_t^{(i)})$ using their majorization/minorization function, i,e, $\nabla_{\bvarphi_t} g_h^{\epsilon}(\bvarphi_t^{(i)})  = \nabla_{\bvarphi_t}v_{h}^{\epsilon}(\bvarphi_t^{(i)})$ and $\nabla_{\bvarphi_t} f(\bvarphi_t^{(i)})  = \nabla_{\bvarphi_t}u(\bvarphi_t^{(i)})$. 

In this regards, substituting \eqref{eq:v_h} in $v_{h,au}^{\epsilon}(\bvarphi_t)$ and $v_{h,cr}^{\epsilon}(\bvarphi_t)$, we have (see \tablename{~\ref{tab:w.r.t vector}} for details),
\begin{equation}
\begin{aligned}
 &v_{h,au}^{\epsilon}(\bvarphi_t) = \textstyle \sum_{k=-N+1}^{N-1} \left(\gamma_{httk}|w_kr_{t,t}(k)|^2 + \mu_{httk} \right) \\
 &=\norm{\bm{\varrho}_{htt}^{\epsilon} \odot (\mathbf{x}_t \circledast \mathbf{x}_t)_k}_2^2 + \textstyle \sum_{k=-N+1}^{N-1} \mu_{httk}
\end{aligned}
\end{equation}
\begin{equation}
\begin{aligned}
 &v_{h,cr}^{\epsilon}(\bvarphi_t) = \textstyle \sum_{\substack{{l=1}\\{l \neq t}}}^{M}\sum_{k=-N+1}^{N-1} \left(\gamma_{htlk}|w_kr_{t,l}(k)|^2 + \mu_{htlk} \right)\\
 &= \textstyle \sum_{\substack{{l=1}\\{l \neq t}}}^{M}\norm{\bm{\varrho}_{htl}^{\epsilon} \odot (\mathbf{x}_l \circledast \mathbf{x}_t)_k}_2^2 + \sum_{\substack{{l=1}\\{l \neq t}}}^{M}\sum_{k=-N+1}^{N-1} \mu_{htlk},
\end{aligned}
\end{equation}
Since, $v_{h}^{\epsilon}(\bvarphi_t) = v_{h,m}^{\epsilon} + v_{h,au}^{\epsilon}(\bvarphi_t) + v_{h,cr}^{\epsilon}(\bvarphi_t)$ and $v_{h,m}^{\epsilon}$ is a constant term, we have,
\begin{equation}\label{eq:dv}
\begin{aligned}
 \nabla_{\bvarphi_t} v_h^{\epsilon}(\bvarphi_t^{(i)})  &= \nabla_{\bvarphi_t}v_{h,au}^{\epsilon}(\bvarphi_t^{(i)}) + \nabla_{\bvarphi_t}v_{h,cr}^{\epsilon}(\bvarphi_t^{(i)}) \\
&= \nabla_{\bvarphi_t}\norm{\bm{\varrho}_{htt}^{\epsilon} \odot (\mathbf{x}_t \circledast \mathbf{x}_t)_k}_2^2\\
&+ \textstyle \sum_{\substack{{l=1}\\{l \neq t}}}^{M}\nabla_{\bvarphi_t}\norm{\bm{\varrho}_{htl}^{\epsilon} \odot (\mathbf{x}_l \circledast \mathbf{x}_t)_k}_2^2,
\end{aligned}
\end{equation}

The gradient of weighted auto- and cross-correlation term term with respect to $\bvarphi_t$ is given by \cite{8168352},
\begin{equation}\label{eq:dv_au}
\begin{aligned}
\nabla_{\bvarphi_t} & \norm{\bm{\varrho}_{htt}^{\epsilon} \odot (\mathbf{x}_t \circledast \mathbf{x}_t)_k}_2^2  =\\
&4\Im\{\mathbf{x}_t^* \odot ((\bm{\varrho}_{htt}^{\epsilon^2} \odot (\mathbf{x}_t \circledast \mathbf{x}_t)) \circledast \mathbf{x}_t )_{k+N-1}\}
\end{aligned}
\end{equation}
and,
\begin{equation}\label{eq:dv_cr}
\begin{aligned}
\nabla_{\bvarphi_t} & \norm{\bm{\varrho}_{htt}^{\epsilon} \odot (\mathbf{x}_m \circledast \mathbf{x}_t)_k}_2^2 =\\
&2\Im\{\mathbf{x}_t^* \odot ((\bm{\varrho}_{htl}^{\epsilon^2} \odot (\mathbf{x}_m \circledast \mathbf{x}_t)^r) \circledast \mathbf{x}_m^* )_{k+N-1}\}
\end{aligned}
\end{equation}
respectively. Substituting \eqref{eq:dv_au} and \eqref{eq:dv_cr} in \eqref{eq:dv}, the gradient $\nabla_{\bvarphi_t} g_h^{\epsilon}(\bvarphi_t^{(i)})$ can be obtained as \eqref{eq:gradient_g}.

On the other hand, \eqref{eq:local approximation Function 2} is a majorizer of $f(\bvarphi_t)$, therefore $\nabla_{\bvarphi_t}f(\bvarphi_t^{(i)})$ is equal to gradient of \eqref{eq:local approximation Function 2} at point $\bvarphi_t^{(i)}$. The second term in \eqref{eq:local approximation Function 2} ($|w_kr_{m,l}(k)|$) is a special case of $g_1(r_{m,l}(k))$, when $p=1$. This term of \eqref{eq:local approximation Function 2} can be majorized by the following equation \cite{7017587}, 
\begin{equation}\label{eq:local approximation Function 4}
\frac{1}{2}|w_k r_{m,l}^{(i)}(k)|^{-1}|w_k r_{m,l}(k)|^2 - \frac{1}{2}|w_k r_{m,l}^{(i)}(k)|.
\end{equation}
Substituting \eqref{eq:local approximation Function 4} with the second term of \eqref{eq:local approximation Function 2}, becomes,
\begin{equation}\label{eq:local approximation Function 5}
	\bar{u}(w_k r_{m,l}(k)) \triangleq \nu_{mlk}|w_k r_{m,l}(k)|^2 + \varsigma_{mlk},
\end{equation}
where,
\begin{equation}
\begin{aligned}
    \nu_{mlk} &\triangleq \frac{p}{2}|w_k r_{m,l}^{(i)}(k)|^{p-2} \\
    \varsigma_{mlk} &\triangleq \eta_{mlk}|w_k r_{m,l}^{(i)}(k)|^2 - \frac{1}{2}\psi_{mlk}|w_k r_{m,l}^{(i)}(k)| \\
    &- (p-1)|w_k r_{m,l}^{(i)}(k)|^p,
\end{aligned}
\end{equation}
and in this case, $\nabla_{\bvarphi_t} f(\bvarphi_t^{(i)}) = \nabla_{\bvarphi_t} \bar{u}(\bvarphi_t^{(i)})$. 

Similar to $v_{h}^{\epsilon}(\bvarphi_t)$, \eqref{eq:local approximation Function 5} can be written as, $\bar{u}(\bvarphi_t) = \bar{u}_m + \bar{u}_{au}(\bvarphi_t) + \bar{u}_{cr}(\bvarphi_t)$, where,
\begin{equation}
\begin{aligned}
 &\bar{u}_{au}(\bvarphi_t) = \textstyle \sum_{k=-N+1}^{N-1} \left(\nu_{ttk}|w_kr_{t,t}(k)|^2 + \varsigma_{ttk} \right) \\
 &=\norm{\bm{\vartheta}_{tt} \odot (\mathbf{x}_t \circledast \mathbf{x}_t)_k}_2^2 + \textstyle \sum_{k=-N+1}^{N-1} \varsigma_{ttk}
\end{aligned}
\end{equation}
\begin{equation}
\begin{aligned}
 &\bar{u}_{cr}(\bvarphi_t) = \textstyle \sum_{\substack{{l=1}\\{l \neq t}}}^{M}\sum_{k=-N+1}^{N-1} \left(\nu_{tlk}|w_kr_{t,l}(k)|^2 + \varsigma_{tlk} \right)\\
 &= \textstyle \sum_{\substack{{l=1}\\{l \neq t}}}^{M}\norm{\bm{\vartheta}_{tl} \odot (\mathbf{x}_l \circledast \mathbf{x}_t)_k}_2^2 + \sum_{\substack{{l=1}\\{l \neq t}}}^{M}\sum_{k=-N+1}^{N-1} \varsigma_{tlk},
\end{aligned}
\end{equation}

Likewise, similar to $\nabla_{\bvarphi_t} g_h^{\epsilon}(\bvarphi_t^{(i)})$, $\nabla_{\bvarphi_t} f(\bvarphi_t^{(i)})$ can be obtained as \eqref{eq:gradient_f}, by replacing, $\bm{\varrho}_{htt}^{\epsilon}$ with $\bm{\vartheta}_{tt}$ and $\bm{\varrho}_{htl}^{\epsilon}$ with $\bm{\vartheta}_{tl}$, respectively.

\subsection{Convergence}
The convergence of proposed method can be discussed in two aspects, the convergence of objective function and the convergence of the waveform set $\bX$. With regard to objective function, as $f(\bX) > 0$ and $g_h^{\epsilon}(\bX) > 0$, therefore, this expression is also valid for the optimum solution of \gls{WeBEST}-e and \gls{WeBEST}-v ($f(\bX^{\star}) > 0$ and $g_h^{\epsilon}(\bX^{\star}) > 0$).

On the other hand, both \gls{WeBEST}-e and \gls{WeBEST}-v minimize the objective function in each step leading to a monotonic decrease of the function value. Since the function value is lower bounded, it can be argued that the algorithm converges to a specific value. Particularly, if the algorithm starts with feasible $\bX^{(0)}$ we have (As well as for $g_h^{\epsilon}(\bX^{\star})$.),
\begin{equation*}
	f(\bX^{(0)}) \geqslant \dots \geqslant f(\bX^{(i)}) \geqslant \dots \geqslant f(\bX^{\star}) > 0,
\end{equation*}


The convergence of the argument requires additional conditions and its investigation is beyond the scope of this paper. However its numerically observed that the argument converges as well as objective function.

\subsection{Computational Complexity}\label{subsec:Computational complexity}
In this subsection we evaluate the computational complexity of \gls{WeBEST}-e and \gls{WeBEST}-v

{\bf Complexity of \gls{WeBEST}-e}: This algorithm needs to perform the following steps in each iteration:
\begin{itemize}
  \item \textit{Calculate the coefficient $u_i$ and $v_i$ in \eqref{eq:Coeff_ui_vi}}: Calculating $u_i$ and $v_i$ needs $M^2N\log_2(N)$ operation due to using fast convolution (see Appendix \ref{app:3} for details). Using a recursive equation, the computational complexity can be reduced to $M N\log_2(N)$.
  \item \textit{Solve the optimization problem 
  \eqref{eq:Pe_v}}: 
  \gls{WeBEST}-e needs finding the roots of $4$ degree polynomials{\footnote{{For finding the roots of polynomial we use ``roots'' function in MATLAB. This function is based on computing the eigenvalues of the companion matrix. Thus the computational complexity of this method is ${\cal{O}}(k^3)$, where $k$ is the degree of the polynomial \cite{1324718, GoluVanl96}}}} in \eqref{eq:dfu_root}, which take $4^3$ operations. In case of discrete phase constraint we obtain \eqref{eq:f_l} using two $L$-points \gls{FFT} which each has $L\log_2(L)$ operations.
  \item \textit{Optimizing all the entries of matrix $\bX$}: To this end we need to repeat the two aforementioned steps $M N$ times.
\end{itemize}
Let us assume that ${\cal{K}}$ iterations are required for convergence of the algorithm. Therefore, the overall computational complexity of 
\gls{WeBEST}-e is ${\cal{O}}({\cal{K}}M N(4^3 + M N\log_2(N)))$, for continuous phase constraint, while under discrete phase constraint is ${\cal{O}}({\cal{K}}M N(L\log_2(L) + M N\log_2(N)))$.

{\bf Complexity of \gls{WeBEST}-v}: This algorithm needs to perform the following steps in each iteration:
\begin{itemize}
    \item \textit{Calculate the gradient of auto- and cross-correlation}: The gradients in \eqref{eq:gradient_f} and \eqref{eq:gradient_g} are expressed in terms of correlations; therefore the gradient needs $N\log_2(N)$ operation due to using fast convolution \cite{8168352}. Since we need to calculate the gradient of auto-correlation for one time and cross-correlation for $M-1$ times, therefore the overall computational complexity would be $MN\log_2(N)$.
    \item \textit{Obtain the step size}: This step contains calculating the auto- and cross-correlation part of objective functions i.e. $f_{au}(\bX)$ and $f_{cr}(\bX)$ ($g_{h,au}^{\epsilon}(\bX)$ and $g_{h,cr}^{\epsilon}(\bX)$), which needs $MN\log_2(N)$ operations. Lets assume that this step needs ${\cal{S}}$ iteration to find the step size, therefore the complexity of this step would be ${\cal{S}}MN\log_2(N)$
    \item \textit{Optimizing all the entries of matrix $\bX$}: To this end we need to repeat the two aforementioned steps $M$ times.
\end{itemize}
Let us assume that ${\cal{K}}$ iterations are required for convergence of the \gls{WeBEST}-v. Therefore, the overall computational complexity of \gls{WeBEST}-v is ${\cal{O}}({\cal{K}}{\cal{S}}M^2N\log_2(N))$.

\section{Numerical Results}\label{sec:Numerical Results}
In this section, we provide representative numerical examples to illustrate the effectiveness of the proposed algorithmic framework. We consider $\Delta\bX^{(i+1)} \triangleq \norm{\bX^{(i+1)} - \bX^{(i)}}_F \leq \zeta$ as the stopping criterion of \gls{WeBEST}-e and \gls{WeBEST}-v, where $\zeta$ is the stopping threshold ($\zeta > 0$). We set $\zeta = 10^{-9}$ for all the following numerical examples. 
We further stop the algorithm if number of iteration exceed $10^5$.
Also, we consider $\epsilon = 0.05$ in \tablename{~\ref{tab:smooth_and_local approximation}}. In this section, by $L\to\infty$ we denote set of continuous phase sequences or set of sequences with infinity alphabet sizes. Besides, we use $10\log(.)$ to report the results based on decibel scale.

\subsection{Convergence}
\begin{figure*}
    \centering
    \begin{subfigure}{.24\textwidth}
        \centering
		\includegraphics[width=1\linewidth]{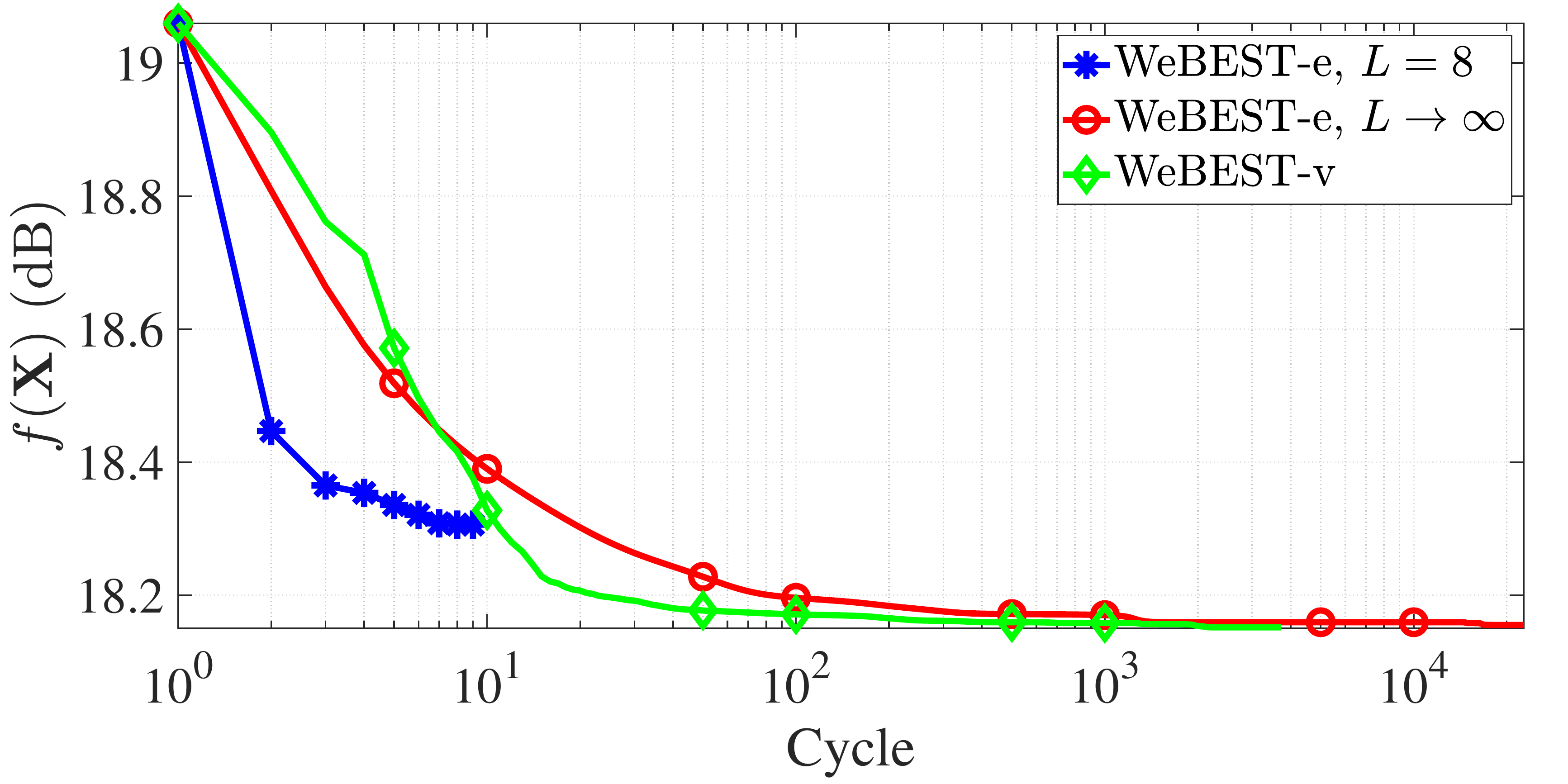}
		\caption[]{$p = 3$}\label{fig:Convergence_fp}
    \end{subfigure}
    \begin{subfigure}{.24\textwidth}
        \centering
		\includegraphics[width=1\linewidth]{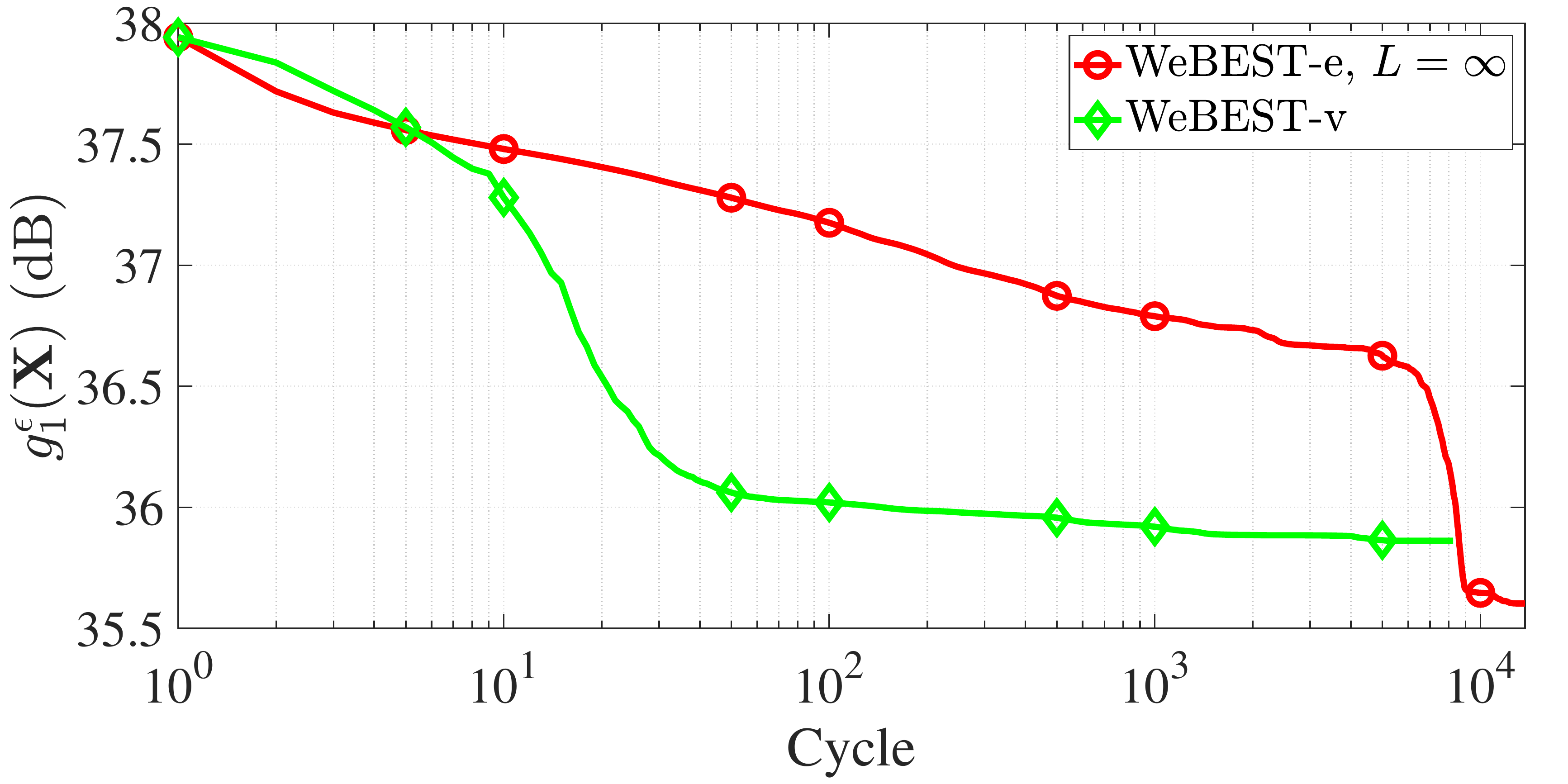}
		\caption[]{$ p = 0.75$, $\epsilon = 0.05$ and $h = 1$}\label{fig:Convergence_g}
    \end{subfigure}
    \begin{subfigure}{.24\textwidth}
        \centering
		\includegraphics[width=1\linewidth]{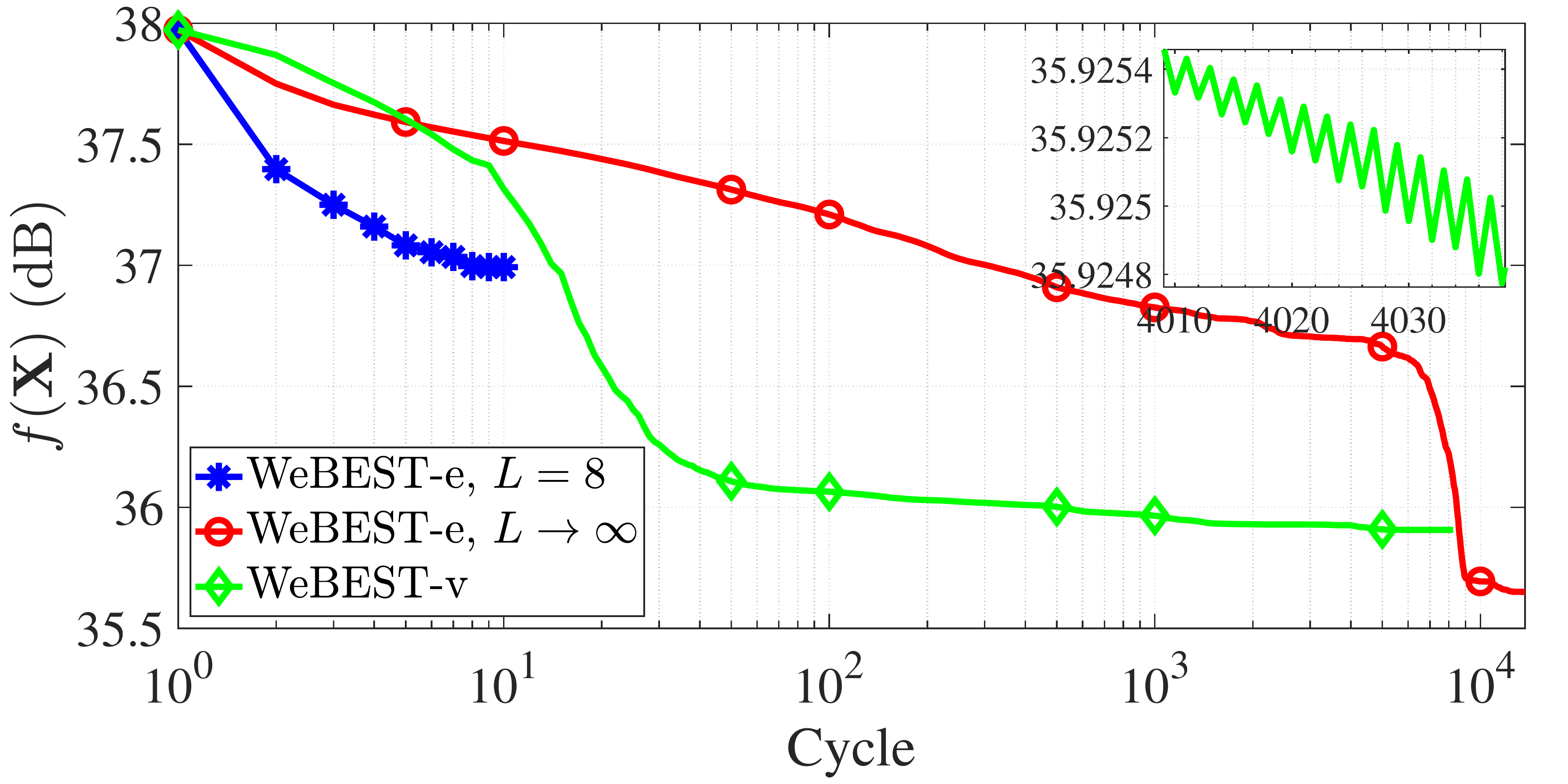}
		\caption[]{The $\ell_p$-norm correspond.}\label{fig:Convergence_fpp}
    \end{subfigure}
    \begin{subfigure}{.24\textwidth}
        \centering
		\includegraphics[width=1\linewidth]{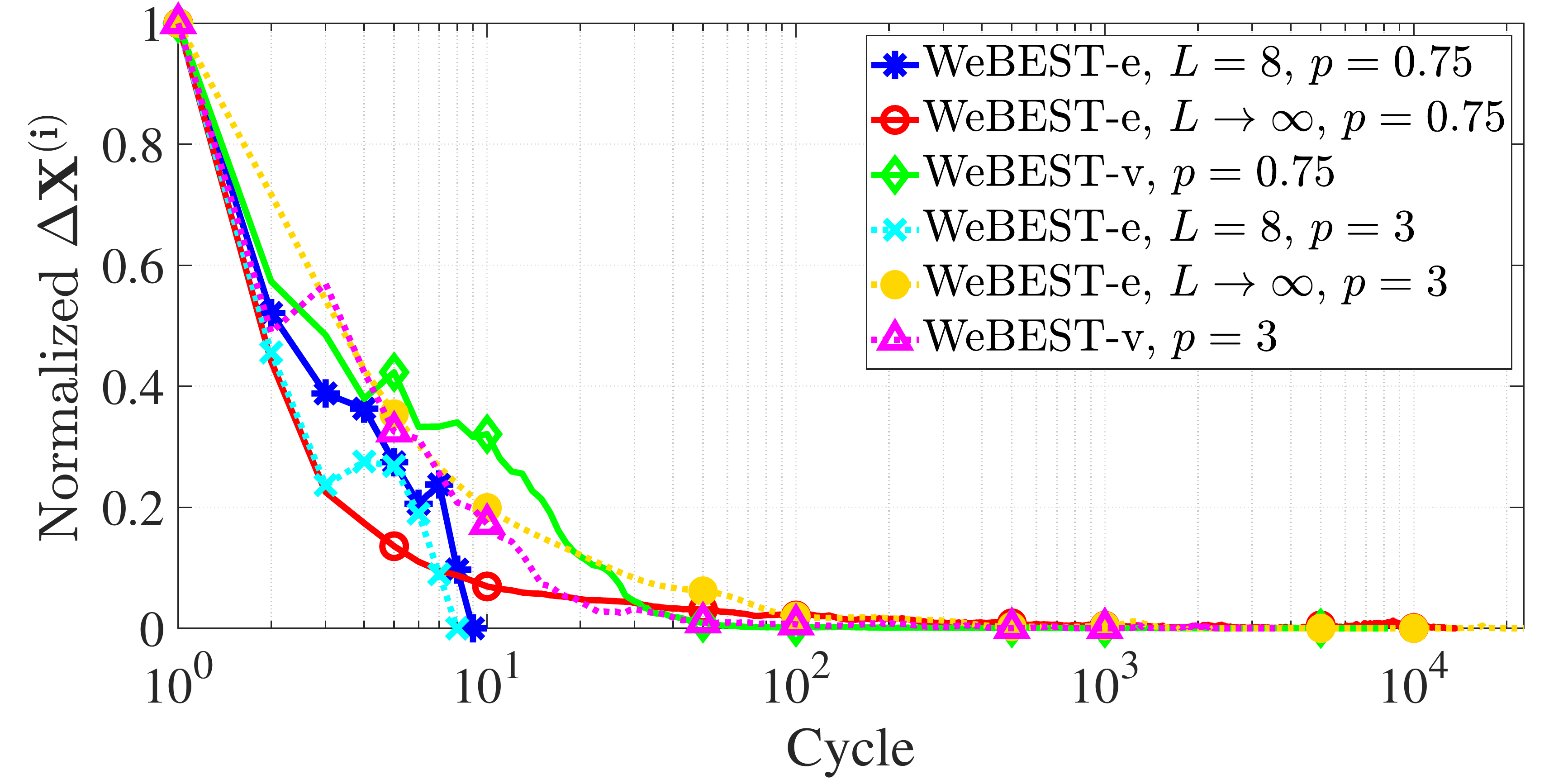}
		\caption[]{Vector Optimization}\label{fig:Convergence_dX}
    \end{subfigure}
    \caption[]{The convergence behavior of proposed method. (a) The $\ell_p$-norm ($f(\bX)$) for $p=3$, (b) the smooth approximation function ($g_1^{\epsilon}(\bX)$) for $p=0.75$, (c) the $\ell_p$-norm correspond to fig (b), and (d) the argument ($\Delta\bX^{(i)}$) ($M=4$ and $N=64$).}\label{fig:Convergence}
\end{figure*}

\figurename{~\ref{fig:Convergence}} depicts the convergence behavior of the proposed method. 
We consider a set of random \gls{MPSK} sequences ($\bX_0 \in \mathbb{C}^{M \times N}$) with number of transmitters $M=4$, code-length $N=64$, and alphabet size $L=8$, as the initial waveform set. For the initialization sequences, every code entry is given by,
\begin{equation}\label{eq:S_0}
	x_{m,n}^{(0)} = e^{j\frac{2\pi(l-1)}{L}},
\end{equation} 
where $l$ is random integer variable uniformly distributed in $[1,L]$. \figurename{~\ref{fig:Convergence_fp}} and {~\ref{fig:Convergence_g}} show the objective function for $p = 3$ ($f(\bX)$) and $p = 0.75$ ($g_1^{\epsilon}(\bX)$) respectively. Observe that, due to the convergence property  of \gls{BSUM} framework, in both cases the objective decreases monotonically. Since for $0 < p \leq 1$ the algorithms is not dealing directly with $\ell_p$-norm metric, the convergence of $f(\bX)$ ($\ell_p$-norm metric) is not monotonic. This fact is shown in \figurename{~\ref{fig:Convergence_fpp}}. However in case of $0 < p \leq 1$, the $f(\bX)$ mimics the monotonous decreasing behavior of the {smooth} approximation function. This shows the accuracy of the {smooth} approximation function. \figurename{~\ref{fig:Convergence_dX}} shows the convergence of the argument when $p = 3$ and $p = 0.75$. 

\subsection{$l_2$-norm (\gls{ISL}) minimization}
In this part we evaluate the performance of proposed method when $p=2$. In this case, the proposed method minimizes the \gls{ISLR} metric ($\text{ISLR} \triangleq \frac{\text{ISL}}{N^2}$) where the lower bound is $10\log(M(M - 1))$ dB \cite{7420715}. \tablename{~\ref{tab:ISLRvsM}} compares the average \gls{ISLR} of the proposed method with Multi-CAN \cite{5072243}, \gls{MM}-Corr \cite{7420715}, \gls{BiST} \cite{8706639} and the lower bound for $N=64$ with different number of transmitters. Similar to the other methods, the proposed method meets the lower bound under continuous phase constraint. Interestingly, in the proposed method even with alphabet size $L= 8$, the obtained set of sequences exhibits the \gls{ISLR} values very close to the lower bound.

\begin{table*}
	\centering
	\caption{Comparison between the \gls{ISLR} (dB) of the proposed method with other methods ($p = 2$, $N=64$).}
	\begin{tabular}[ht]
	    {c|c|c|c|c|c|c|c|c|c}
		\hline
		\hline
		$M$ & 2 & 3 & 4 & 5 & 6 & 7 & 8 & 9 & 10  \\
		\hline
		Initial & 5.9289 & 9.8565 & 11.9106 & 14.0384 & 15.5558 & 16.8349 & 18.0590 & 19.2051 & 19.9744 \\ \hline
		Lower bound & 3.0103 & 7.7815 & 10.7918 & 13.0103 & 14.7712 & 16.2325 & 17.4819 & 18.5733 & 19.5424 \\ \hline
		\gls{WeBEST}-e, $L\to\infty$ & 3.0103 & 7.7815 & 10.7918 & 13.0103 & 14.7712 & 16.2325 & 17.4819 & 18.5733 & 19.5424 \\\hline
		\gls{WeBEST}-v & 3.0103 & 7.7815 & 10.7918 & 13.0103 & 14.7712 & 16.2325 & 17.4819 & 18.5733 & 19.5424 \\ \hline
		Multi-CAN & 3.0103 & 7.7815 & 10.7918 & 13.0103 & 14.7712 & 16.2325 & 17.4819 & 18.5733 & 19.5424 \\\hline
		\gls{MM}-Corr & 3.0103 & 7.7815 & 10.7918 & 13.0103 & 14.7712 & 16.2325 & 17.4819 & 18.5733 & 19.5424 \\\hline
		\gls{WeBEST}-e, $L=8$ & 3.2582 & 7.8695 & 10.8284 & 13.0319 & 14.7840 & 16.2404 & 17.4888 & 18.5779 & 19.5463 \\\hline
		\gls{BiST} ($\theta = 0$, $L=8$) & 3.2632 & 7.8529 & 10.8238 & 13.0302 & 14.7901 & 16.2411 & 17.4884 & 18.5796 & 19.5458 \\\hline
		\hline
	\end{tabular}
	\label{tab:ISLRvsM}
\end{table*}

\tablename{~\ref{tab:ISLRvsN}} shows the optimized \gls{ISLR} values under discrete phase constraint, for different sequence lengths when $M = 4$. In this table, we consider to assess the performance of the proposed method with alphabet size of $L=8$. Referring to the lower bound in the \tablename{~\ref{tab:ISLRvsM}}, we observe that the optimized sequences have \gls{ISLR} values quite close to the lower bound.


\begin{table}
	\centering
	\caption{The \gls{ISLR} obtained by the proposed method under discrete phase constraint with different length ($p = 2$, $M=4$).}
	\begin{tabular}{c|c|c|c|c|c}	
		\hline
		\hline
		$N$ & 64 & 128  & 256  & 512  & 1024 \\
		\hline
		$L=8$ & 10.8245 & 10.8253 & 10.8251 & 10.8220 & 10.8237  \\
		\hline
		\hline
	\end{tabular}
	\label{tab:ISLRvsN}
\end{table}

\subsection{$\ell_p$-norm minimization for $p>2$}
Best \gls{PSL} values can be obtained by $\ell_p$-norm minimization of the auto- and cross-correlation, when $p\to\infty$. To this end, we consider a increasing scheme for selection of $p$ in several steps. Specifically, we consider the $p$ steps as, $2 \leq p_1 < p_2 < \dots < p_T < \infty$. Particularly, we start with a random set of sequences as initial waveform and we optimize the $\ell_{p_1}$-norm of auto- and cross-correlation functions. Then we select the optimized solution of $\ell_{p_1}$-norm as the initial waveform for $l_{p_2}$ minimization. Subsequently we repeat this procedure until we cover all of the $p_i$ values ($i\in{1,\dots,T}$).

\begin{figure}
	\centering
	\includegraphics[width=1.0\columnwidth]{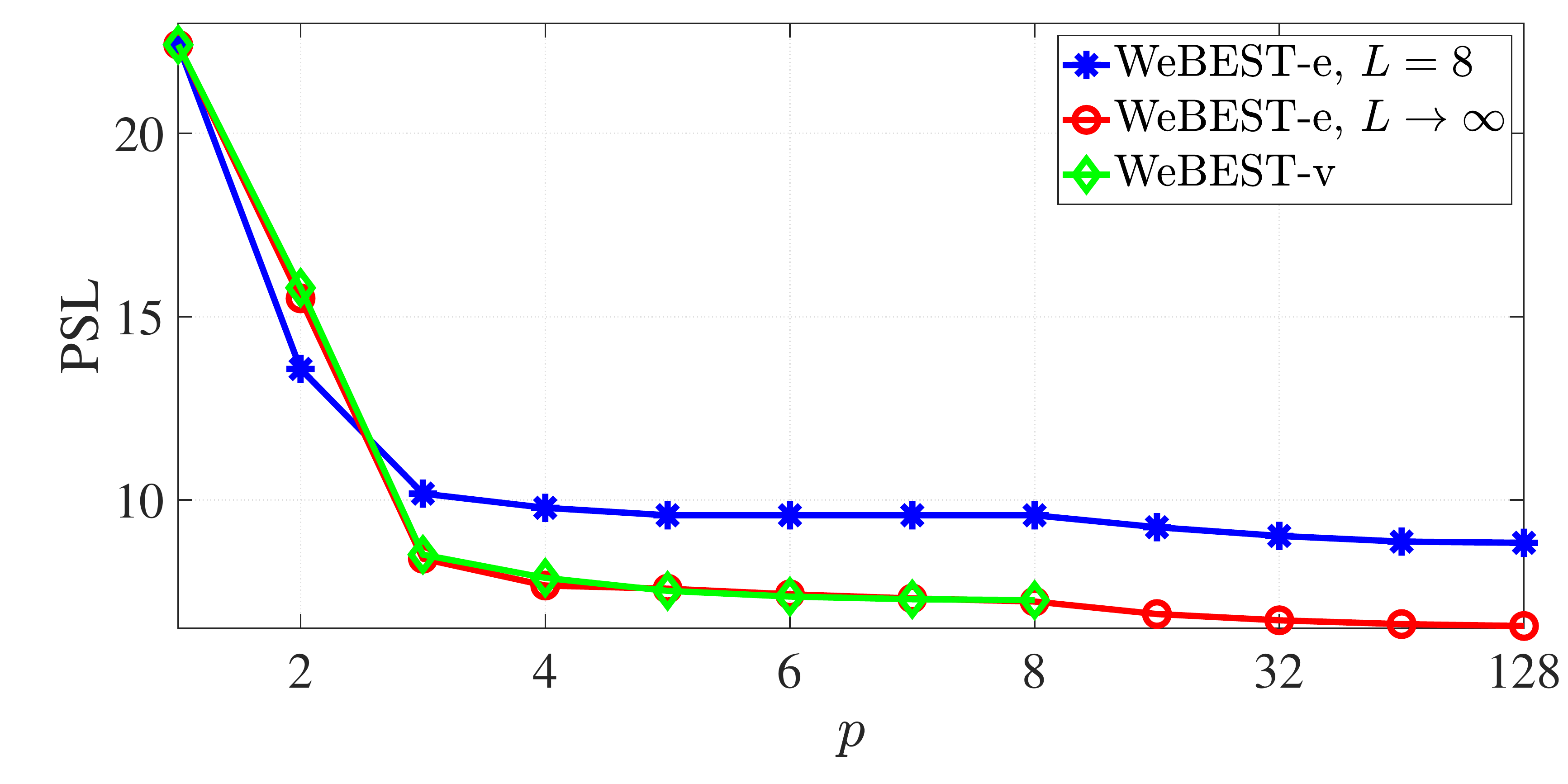}
	\caption{The \gls{PSL} behavior vs $p$. ($M=4$ and $N = 64$)}
	\label{fig:PSL_vs_p}
\end{figure}

\figurename{~\ref{fig:PSL_vs_p}} shows the performance of \gls{PSL} minimization of the proposed method based on aforementioned approach. In this figure we assume that both \textbf{Algorithm \ref{alg:vector_optimization}} and \textbf{Algorithm \ref{alg:entry_optimization}} are initialized with the same random \gls{MPSK} sequence with $L=8$. As can be seen from \figurename{~\ref{fig:PSL_vs_p}}, the \gls{PSL} decreases and converge to the optimum \gls{PSL} for vector and entry optimization under discrete and continuous phase constraints\footnote{Please note that to obtain results for large $p$ values a normalization for the objective is required due to the numerical issues. In this paper, we report the results without performing any normalization of the objective, for $p \leq 128$ and $p \leq 8$ 
in cases of entry and vector optimization, respectively.}.

\begin{figure*}
    \centering
    \begin{subfigure}{.49\textwidth}
        \centering
		\includegraphics[width=1\linewidth]{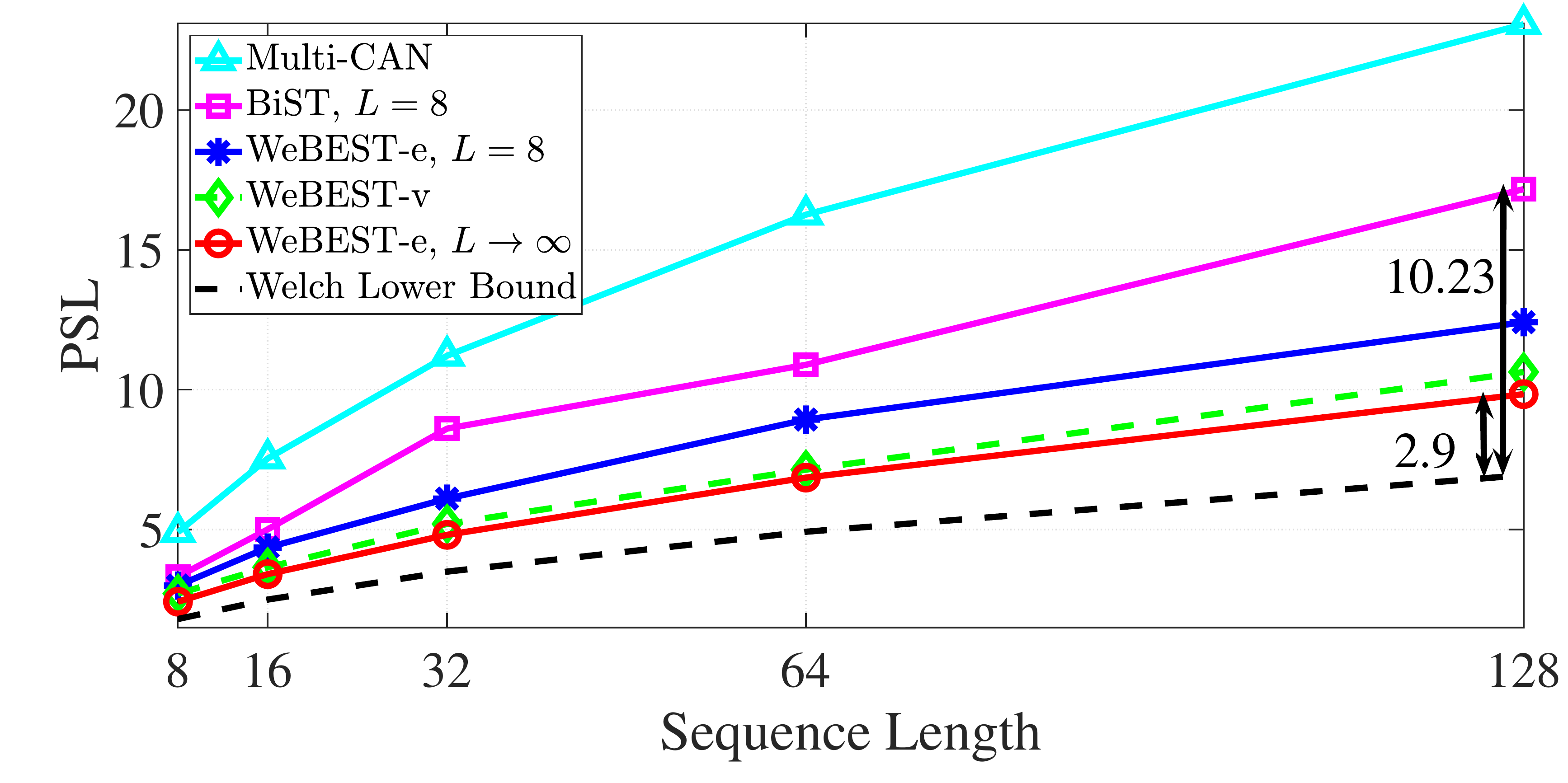}
		\caption[]{\gls{PSL} versus sequence length ($M = 4$).}\label{fig:PSL_vs_N}
    \end{subfigure}
    \begin{subfigure}{.49\textwidth}
        \centering
		\includegraphics[width=1\linewidth]{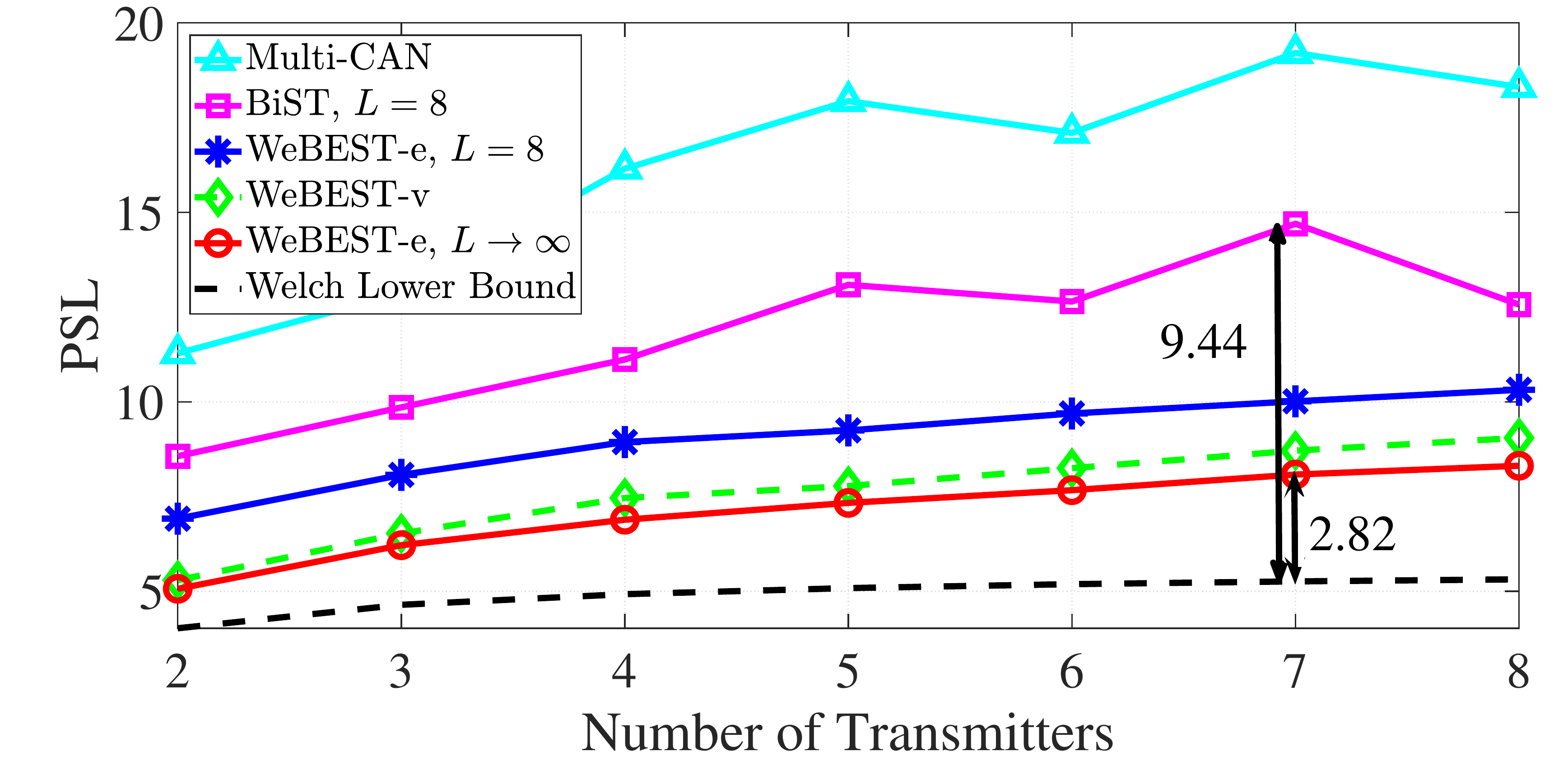}
		\caption[]{\gls{PSL} versus antenna number ($N=64$).}\label{fig:PSL_vs_M}
    \end{subfigure}
    \caption[]{Comparing the performance of the proposed method with Multi-\gls{CAN}, \gls{BiST} and Welch lower bound in terms of \gls{PSL}.}\label{fig:PSL_vs_pNM}
\end{figure*}

In \figurename{~\ref{fig:PSL_vs_N}}, we fixed the number of transmitters ($M=4$) and report the \gls{PSL} with different sequence length.Vice versa in \figurename{~\ref{fig:PSL_vs_M}}, we fixed the sequence length $N=64$ and report the \gls{PSL} with different number of transmitters. In both figures we compare the performance of proposed method with \gls{BiST} \cite{8706639} in \gls{PSL} minimization mode ($\theta = 1$), Multi-\gls{CAN} \cite{5072243} and the Welch lower band for \gls{PSL}, which is \cite{9781139095174},
\begin{equation}
    B_{PSL} = \sqrt{\frac{M-1}{2MN-M-1}}.
\end{equation}

As can be seen, the \gls{WeBEST} obtains lower \gls{PSL} values, i.e., closer to the Welch lower bound when compared with its counterparts. Indeed, \gls{WeBEST} decreases the gap between the \gls{PSL}'s obtained by the state of the art with the Welch lower bound significantly. For instance in \figurename{~\ref{fig:PSL_vs_N}} for $N=128$ the gap between \gls{WeBEST}-e ($L\to\infty$) and Welch is about $2.9$, whereas this gap for \gls{BiST} is about $10.23$.

\subsection{$\ell_p$-norm minimization for $0 < p \leq 1$}


Obtaining a sparse auto- and cross-correlation is equivalent with minimizing the $\ell_p$-norm of the auto- and cross-correlation, when $p \to 0$. To develop $\ell_p$-norm minimization, inverse to \gls{PSL} minimization we consider decreasing the value of $p$ in several steps, where $1 \geq p_1 > p_2 > \dots > p_T > 0$. In order to evaluate the performance of $l_0$-norm minimization we consider a threshold for the lags of auto- and cross-correlations. If the absolute value of the lags is less than that threshold, we assume that the lags is zero. In constant modulus sequence, since $|r_{m,l}(N-1)| = |x_{m,N}x_{l,1}^*| = 1$, the lowest possible \gls{PSL} is equal to $1$ \cite{iet_ch3}. Therefore, we chose $1$ as the threshold.
Let $N_s$ be the numbers of lags of auto- and cross-correlation which their absolute value is less than $1$. We introduce the sparsity as,
\begin{equation*}
	S_p = \frac{N_s}{M^2(2N-1)}
\end{equation*}
where, the denominator ($M^2(2N-1)$) is the total number of lags of auto and cross-correlations. $S_p \in [0, 1]$ and if $S_p \to 1$ means the auto- and cross-correlation of set of sequence is sparse and vice versa if $S_p \to 0$ means the auto- and cross-correlation of set of sequence is not sparse. 

\begin{figure*}
    \centering
    \begin{subfigure}{.32\textwidth}
        \centering
		\includegraphics[width=1\linewidth]{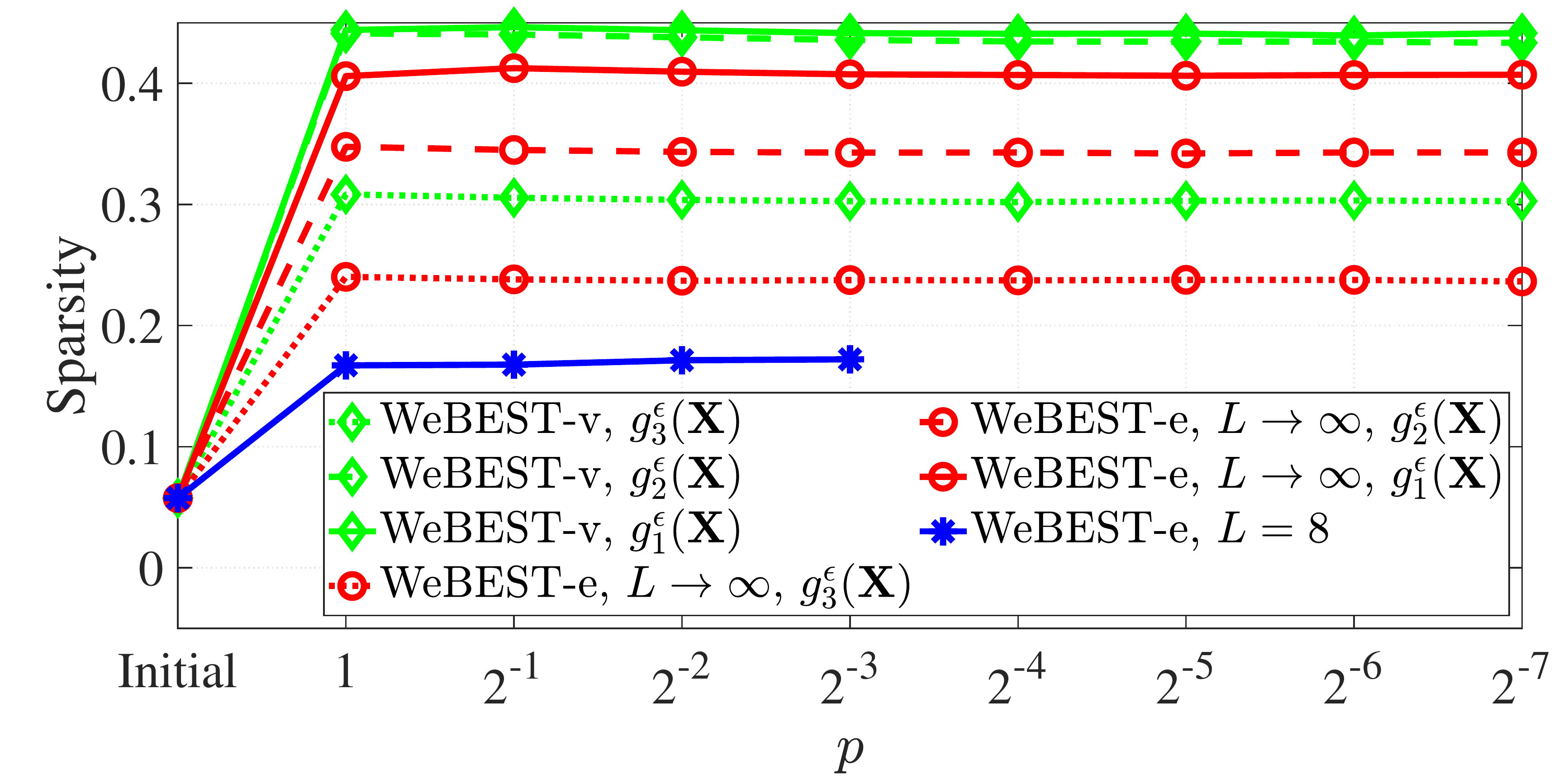}
		\caption[]{Sparsity versus $p$ ($M=4$ and $N=64$).}\label{fig:L0_vs_p}
    \end{subfigure}
    \begin{subfigure}{.32\textwidth}
        \centering
		\includegraphics[width=1\linewidth]{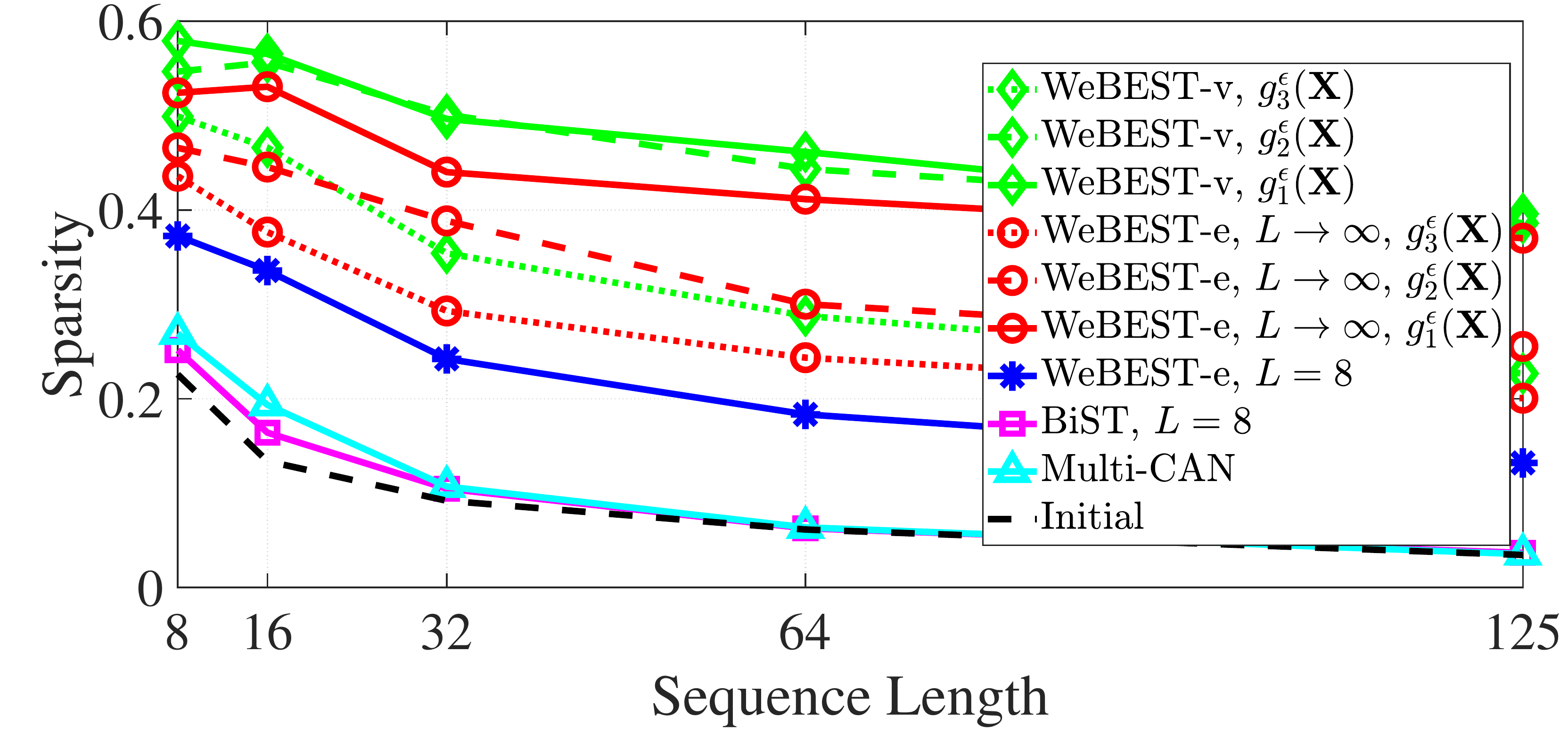}
		\caption[]{Sparsity versus sequence length ($M = 4$).}\label{fig:L0_vs_N}
    \end{subfigure}
    \begin{subfigure}{.32\textwidth}
        \centering
		\includegraphics[width=1\linewidth]{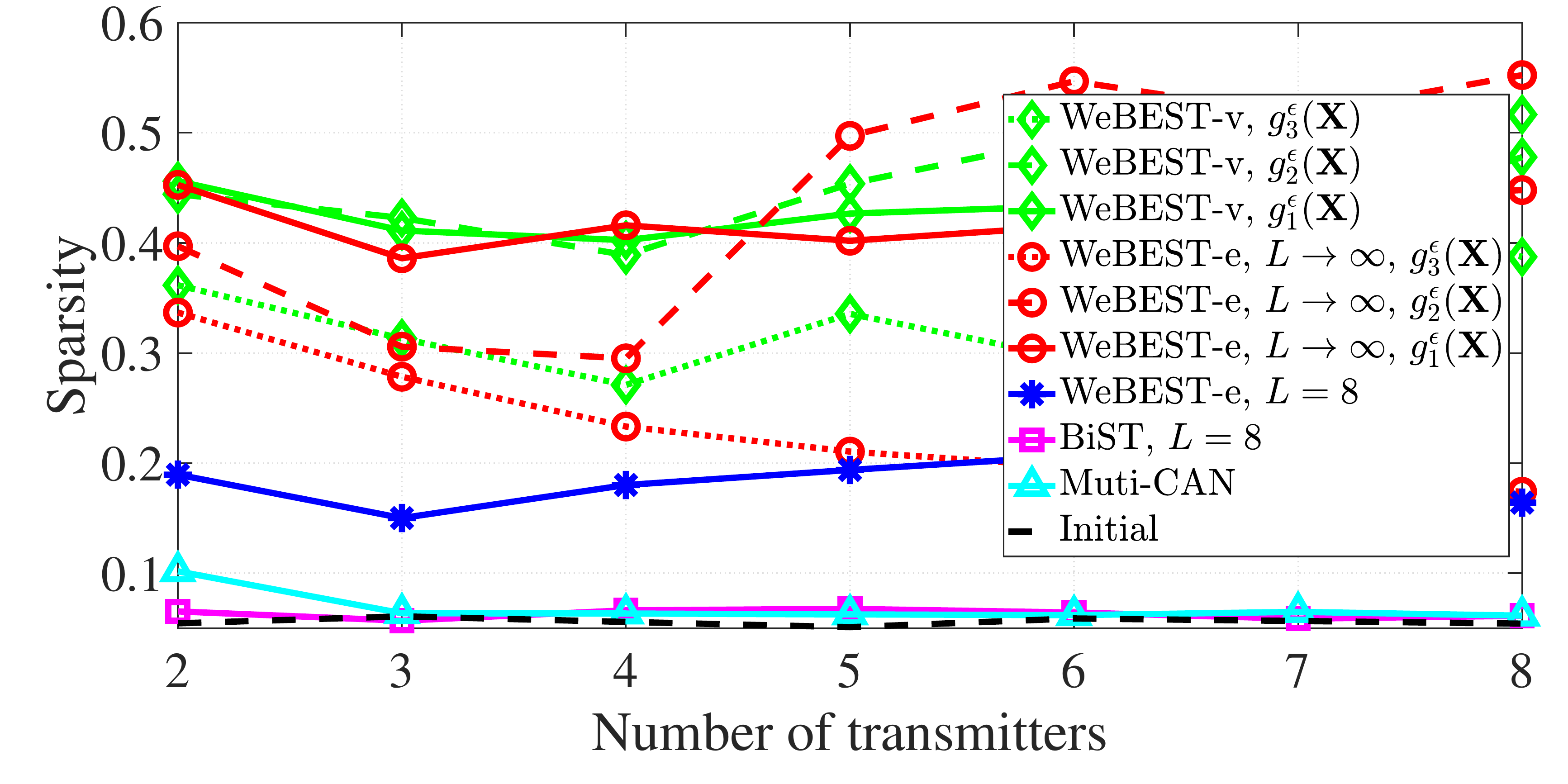}
		\caption[]{Sparsity versus transmitter ($N=64$).}\label{fig:L0_vs_M}
    \end{subfigure}
    \caption[]{The Sparsity behavior and comparing the performance of the proposed method with other methods.}\label{fig:L0_vs_pNM}
\end{figure*}

\figurename{~\ref{fig:L0_vs_pNM}} shows the sparsity obtained by the proposed method based on aforementioned approach. In this figure, we initialize  both \textbf{Algorithm \ref{alg:vector_optimization}} and \textbf{Algorithm \ref{alg:entry_optimization}} with identical random \gls{MPSK} sequence with $L=8$. As can be seen from \figurename{~\ref{fig:L0_vs_p}}, the sparsity increases and converges to the optimum value for vector and entry optimizations under discrete and continuous phase constraints. In \figurename{~\ref{fig:L0_vs_N}} and \figurename{~\ref{fig:PSL_vs_M}}, we evaluate the sparsity obtained by \gls{WeBEST} when the number of the antenna is fixed at $M=4$ with different sequence lengths and vice versa when the sequence length is fixed at $N=64$ with different number of transmitters. In both figures we compare the performance of proposed method with \gls{BiST} \cite{8706639} in \gls{ISL} minimization mode ($\theta = 0$) and Multi-CAN. As can be seen the proposed method obtains higher sparsity when compared with its counterparts. 

\subsection{The impact of $p$}
Here we evaluate the impact of $p$ on the auto- and cross-correlations. \figurename{~\ref{fig:L0_vs_pNM}} shows the auto-correlation of the first sequence with three different values of $p$ namely, $p \to 0$, $p = 2$ and $p \to \infty$ for entry and vector optimization procedures under discrete phase constraint. As can be seen in all the cases when $p \to 0$, the auto-correlation function has many lags which are below the sparsity threshold. When $p=2$, the proposed method offers a waveform with good \gls{ISL} property. By increasing the $p \to \infty$ the lags become flat and the algorithm offers a waveform with good \gls{PSL} property.

\begin{figure*}
    \centering
    \begin{subfigure}{.32\textwidth}
        \centering
		\includegraphics[width=1\linewidth]{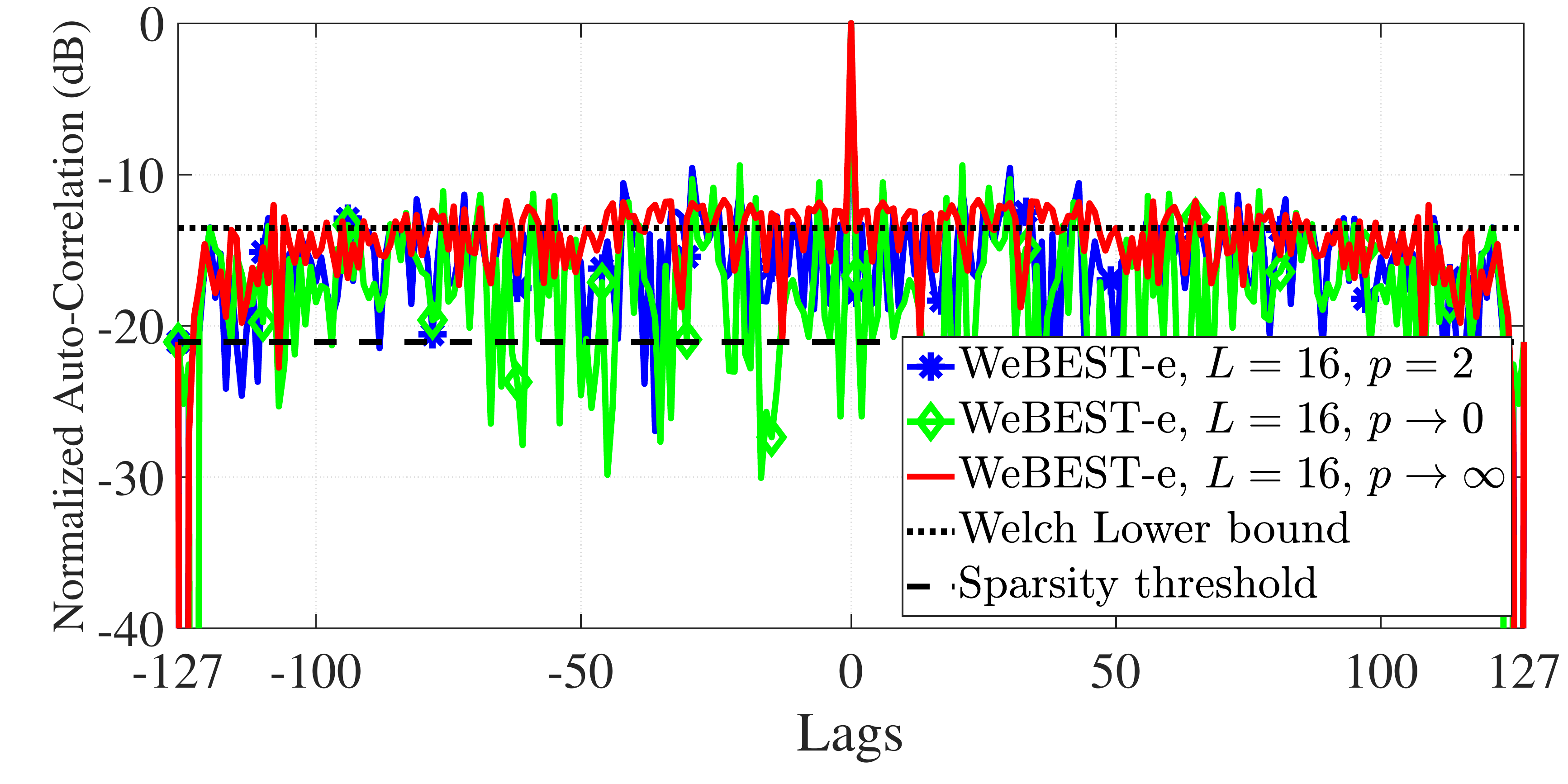}
		\caption[]{Discrete phase ($L=16$).}\label{fig:ACd_vs_p}
    \end{subfigure}
    \begin{subfigure}{.32\textwidth}
        \centering
		\includegraphics[width=1\linewidth]{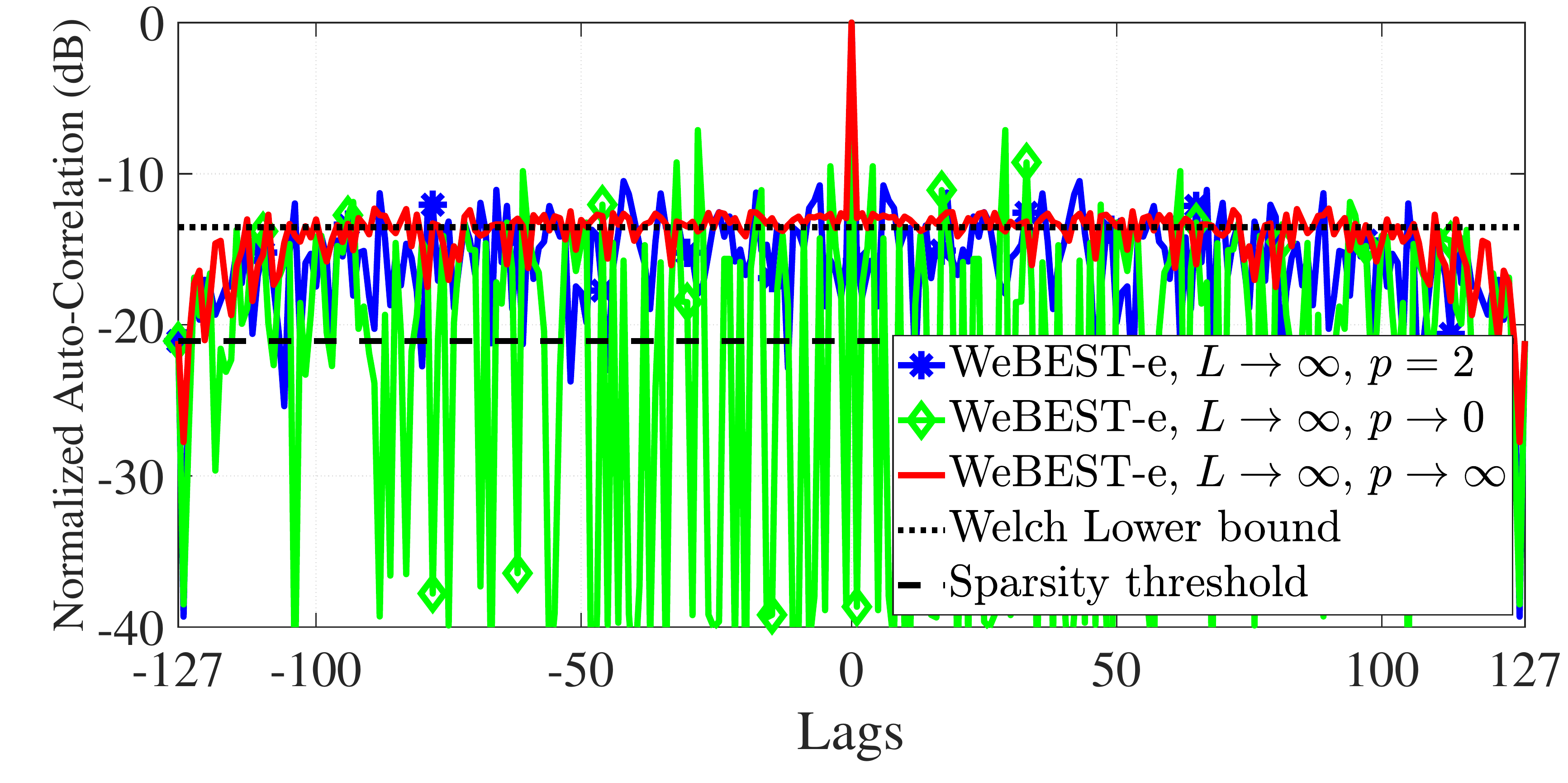}
		\caption[]{Entry optimization.}\label{fig:ACe_vs_p}
    \end{subfigure}
    \begin{subfigure}{.32\textwidth}
        \centering
		\includegraphics[width=1\linewidth]{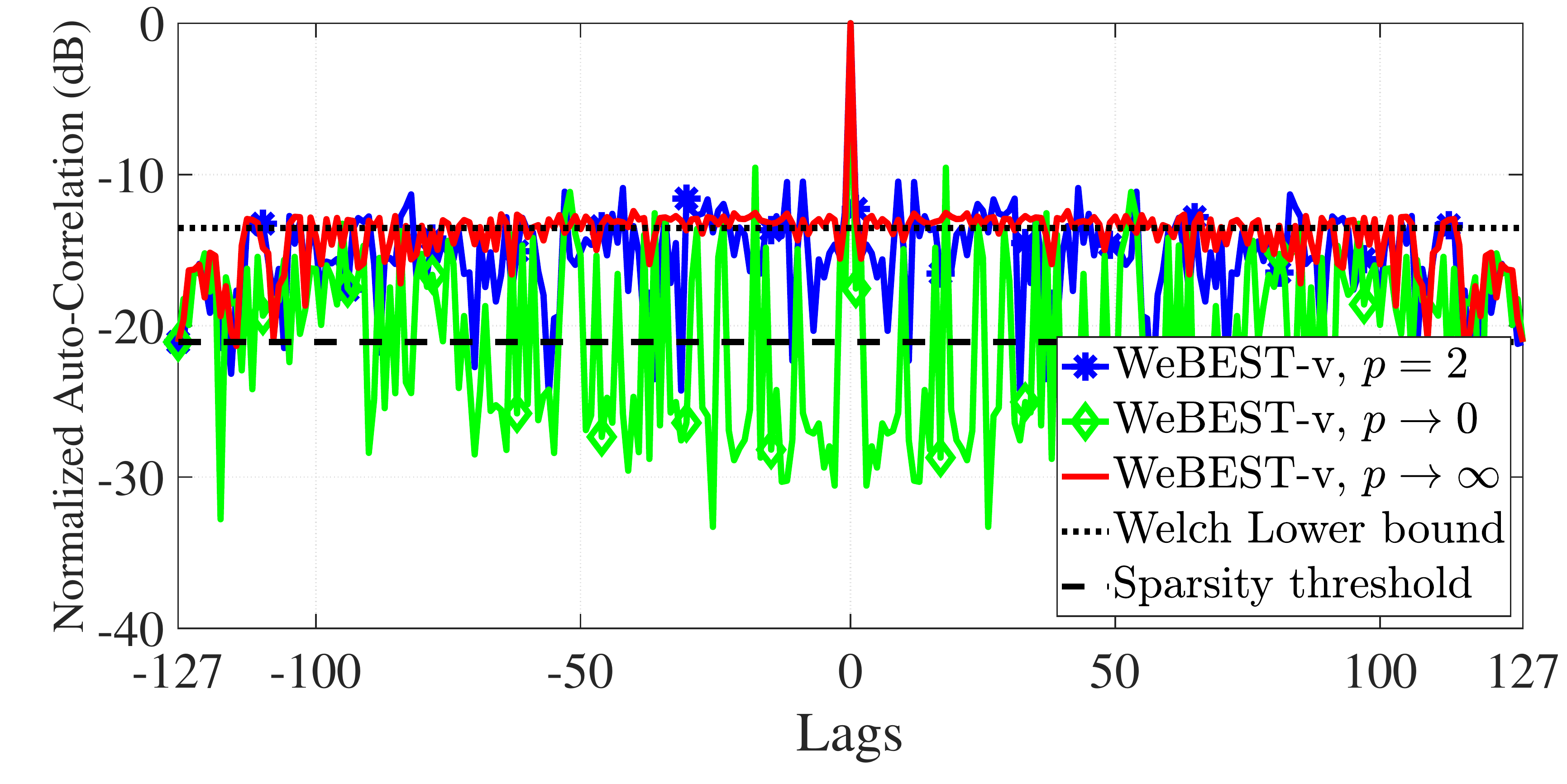}
		\caption[]{Vector optimization.}\label{fig:ACv_vs_p}
    \end{subfigure}
    \caption[]{The impact of choosing $p$ on auto-correlation ($M=2$ and $N=128$). }\label{fig:AC_vs_p}
\end{figure*}

\subsection{The impact of weighting ($\mathbf{w}$)}
In this part we evaluate the impact of weighting on the auto- and cross-correlation of the proposed method. Let $\mathcal{V}$ and $\mathcal{U}$ be the desired and undesired lags for \gls{MIMO} radar, respectively. These two sets satisfy $\mathcal{V} \cup \mathcal{U} = \{-N+1,\dots,N-1\}$ and $\mathcal{V} \cap \mathcal{U} = \emptyset$. We assume that,
$$\begin{dcases}
	w_k = 1, & k \in \mathcal{V}\\
	w_k = 0, & k \in \mathcal{U}
\end{dcases}$$

\begin{figure*}
    \centering
    \begin{subfigure}{.32\textwidth}
        \centering
		\includegraphics[width=1\linewidth]{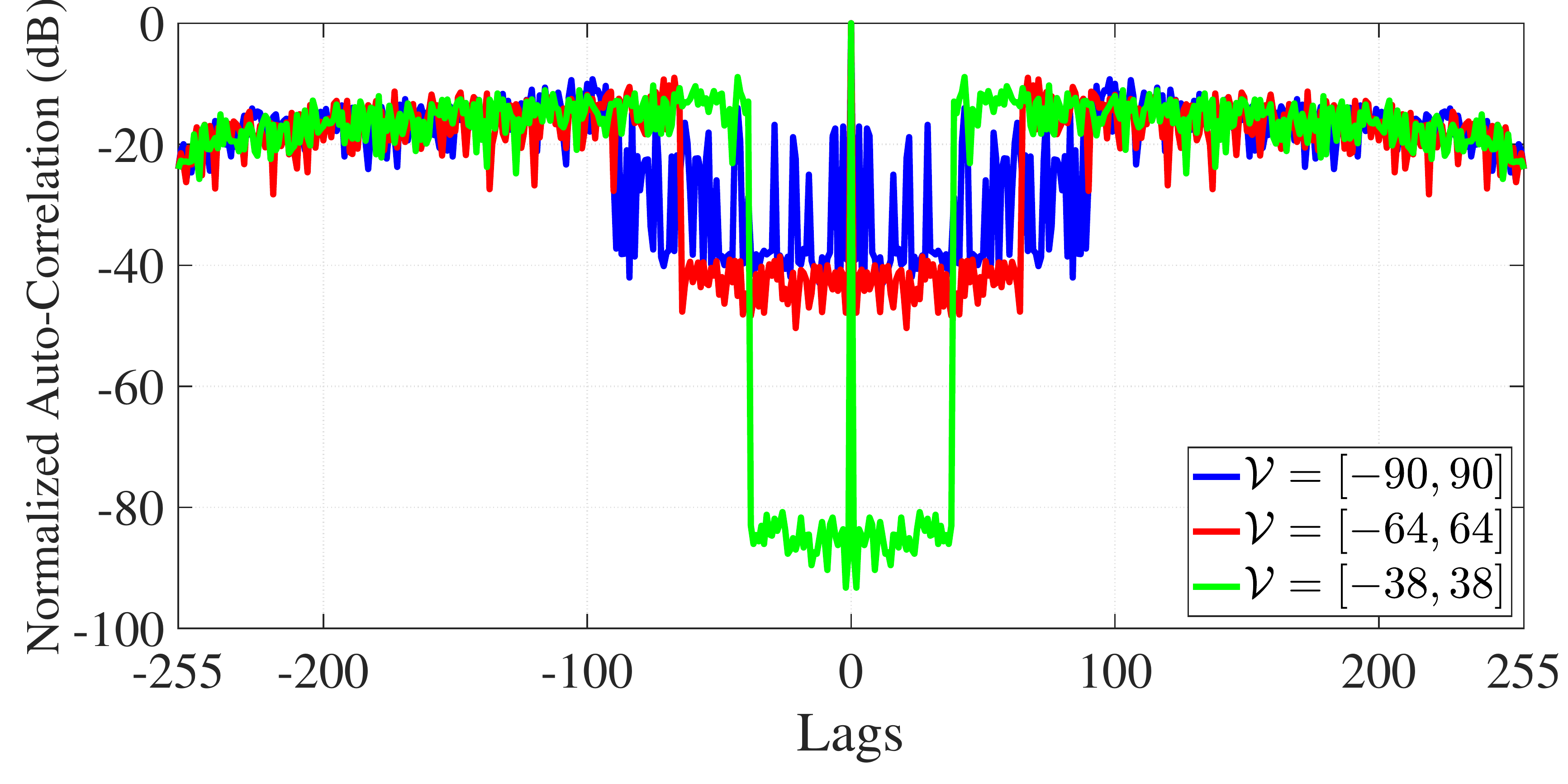}
		\caption[]{$p\to0$.}\label{fig:ACd_vs_wp}
    \end{subfigure}
    \begin{subfigure}{.32\textwidth}
        \centering
		\includegraphics[width=1\linewidth]{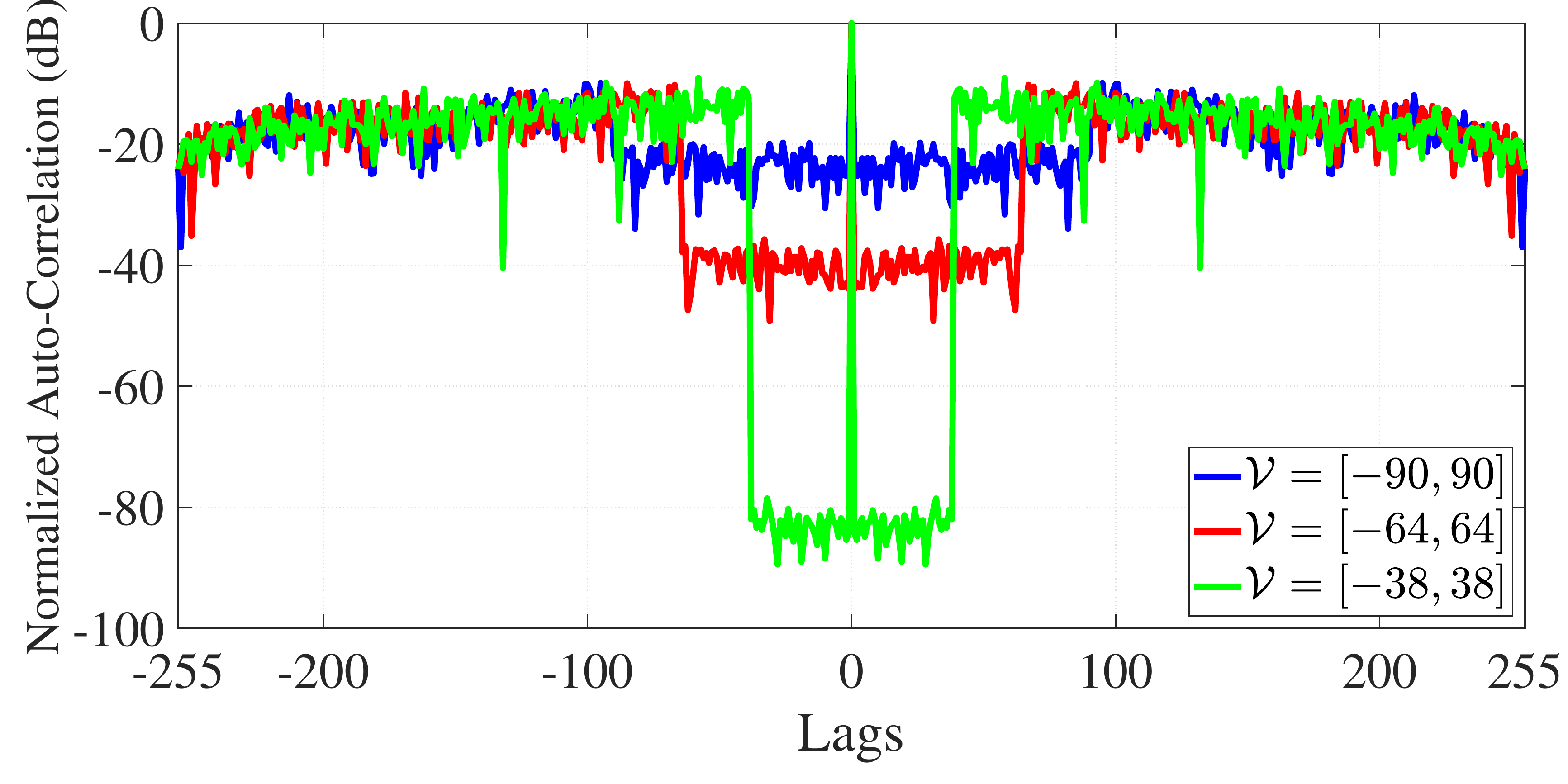}
		\caption[]{$p=2$.} \label{fig:ACv_vs_wp}
    \end{subfigure}
    \begin{subfigure}{.32\textwidth}
        \centering
		\includegraphics[width=1\linewidth]{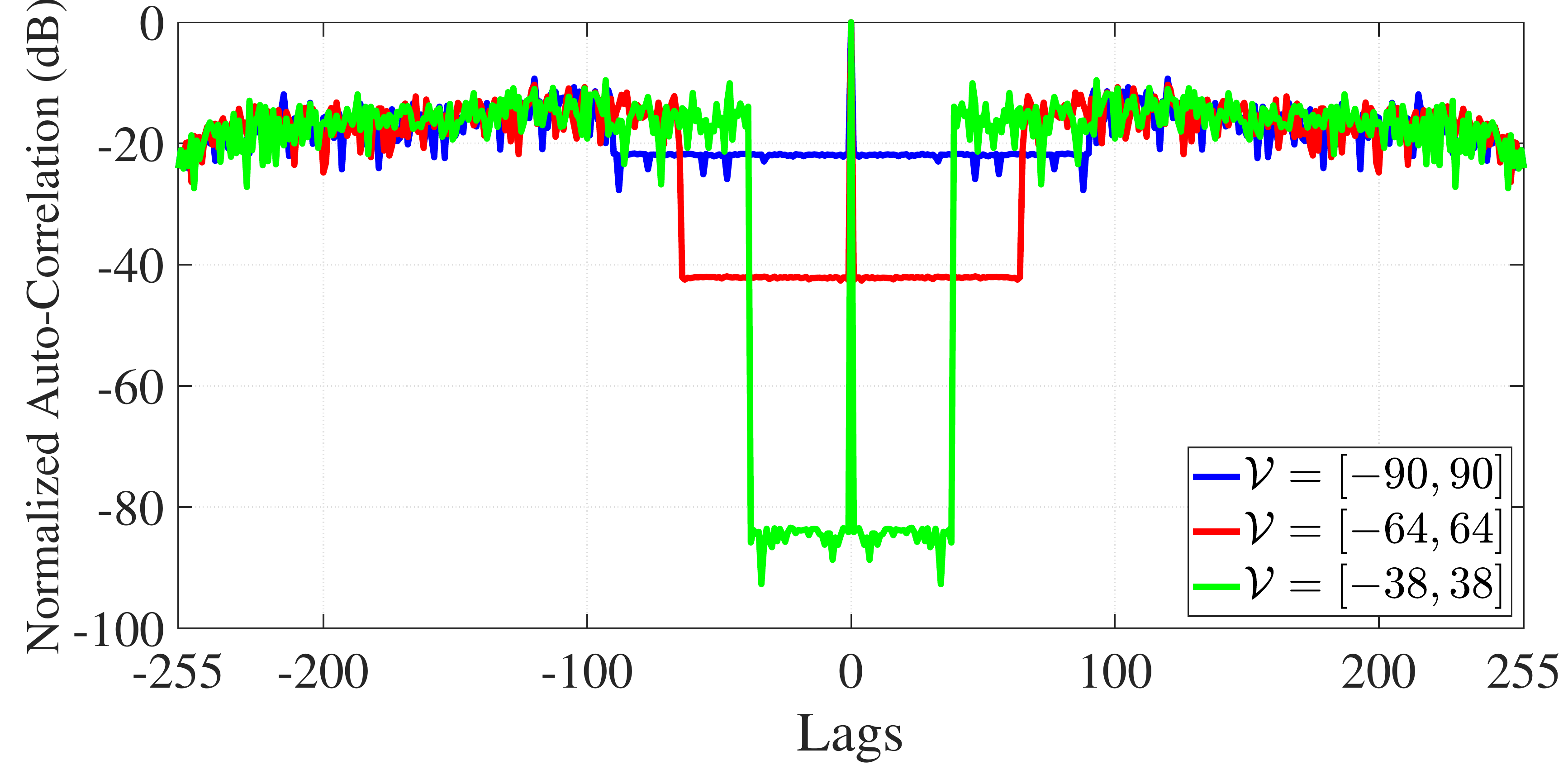}
		\caption[]{$p \to \infty$.}\label{fig:ACe_vs_wp}
    \end{subfigure}
    \caption[]{The impact of weighting in \gls{WeBEST}-e with different values of $p$ ($L\to\infty$, $M=2$ and $N=256$).}\label{fig:AC_vs_wp}
\end{figure*}

\figurename{~\ref{fig:AC_vs_wp}} shows the impact of weighting with $M=2$, $N=256$ and different values of $p$, under continuous phase and entry-based optimization. In addition, we assume different region of desired lags, namely, $\mathcal{V} = [-90, 90]$, $\mathcal{V} = [-64, 64]$ and $\mathcal{V} = [-38, 38]$. As can be seen, by decreasing the range we obtain a deeper null and vice versa in all cases. Besides, in \figurename{~\ref{fig:ACd_vs_wp}}, \figurename{~\ref{fig:ACe_vs_wp}} and \figurename{~\ref{fig:ACv_vs_wp}} we obtain sparse, good \gls{PSL} and good \gls{ISL} of auto-correlation in desired regions of lags.



\begin{figure*}
    \centering
    \begin{subfigure}{.32\textwidth}
        \centering
		\includegraphics[width=1\linewidth]{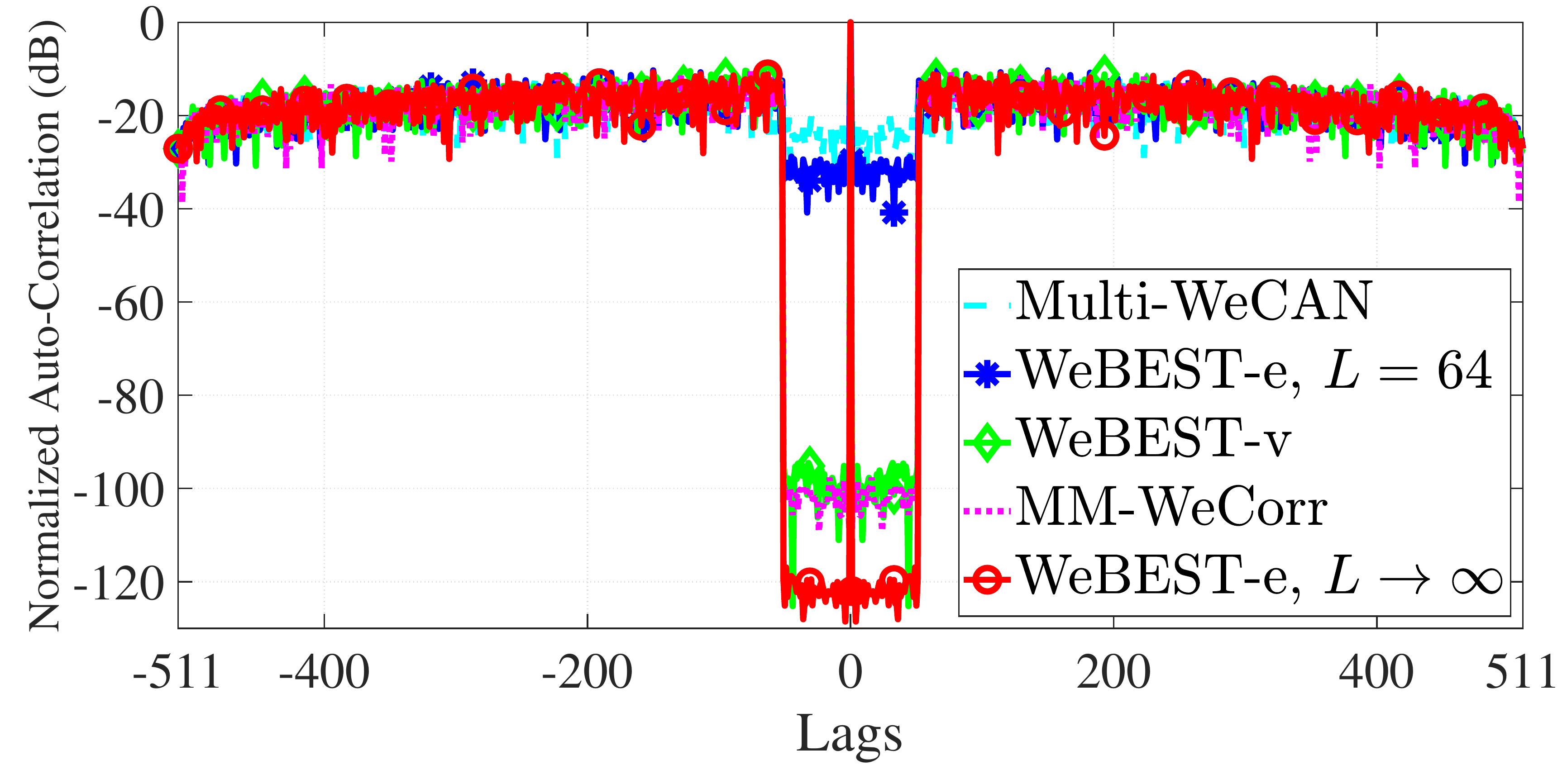}
		\caption[]{The Auto-Correlation of the first waveform.}\label{fig:AC1_vs_w}
    \end{subfigure}
    \begin{subfigure}{.32\textwidth}
        \centering
		\includegraphics[width=1\linewidth]{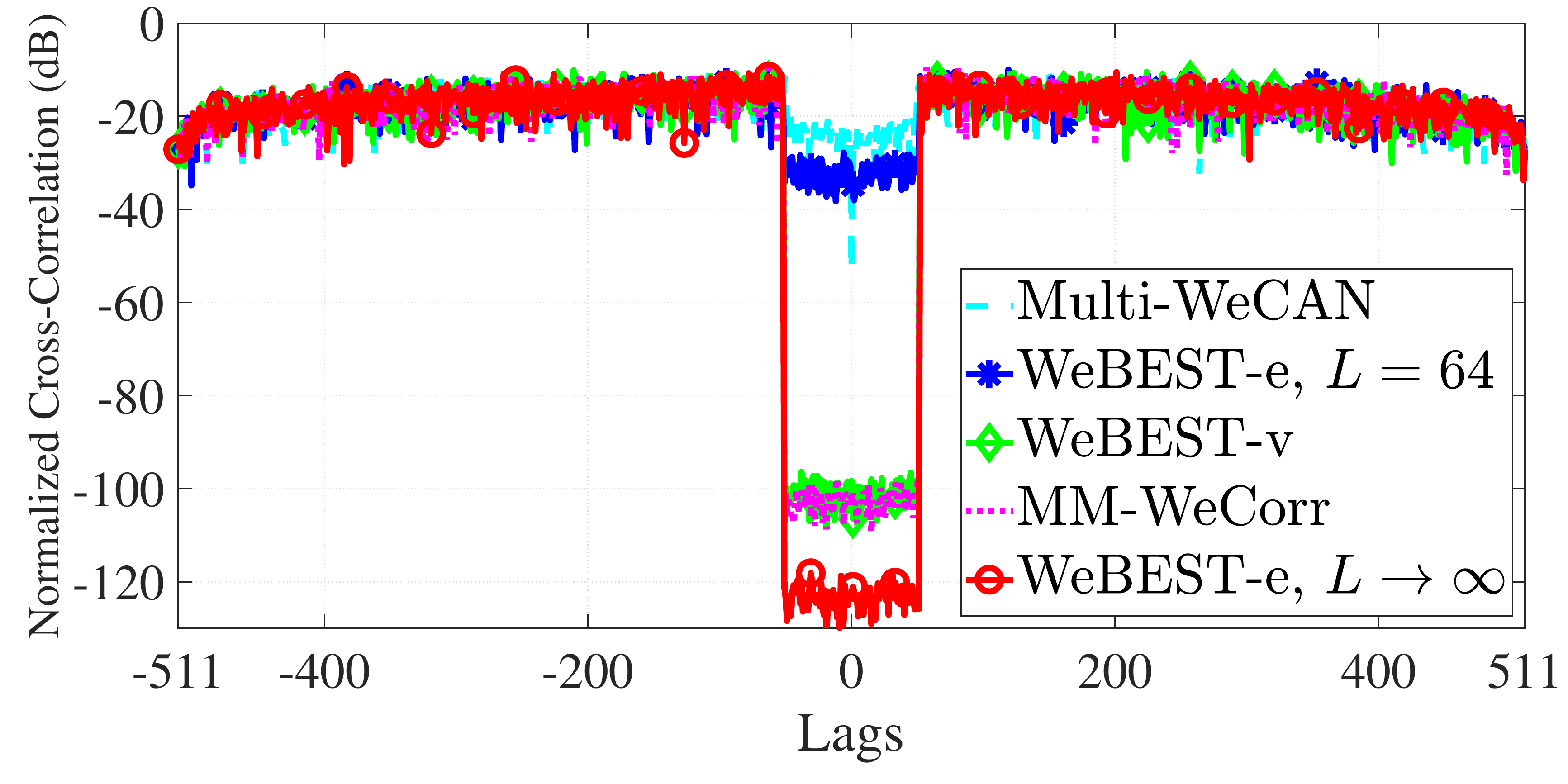}
		\caption[]{The Cross-Correlation of the first and second waveforms.}\label{fig:C12_vs_w}
    \end{subfigure}
    \begin{subfigure}{.32\textwidth}
        \centering
		\includegraphics[width=1\linewidth]{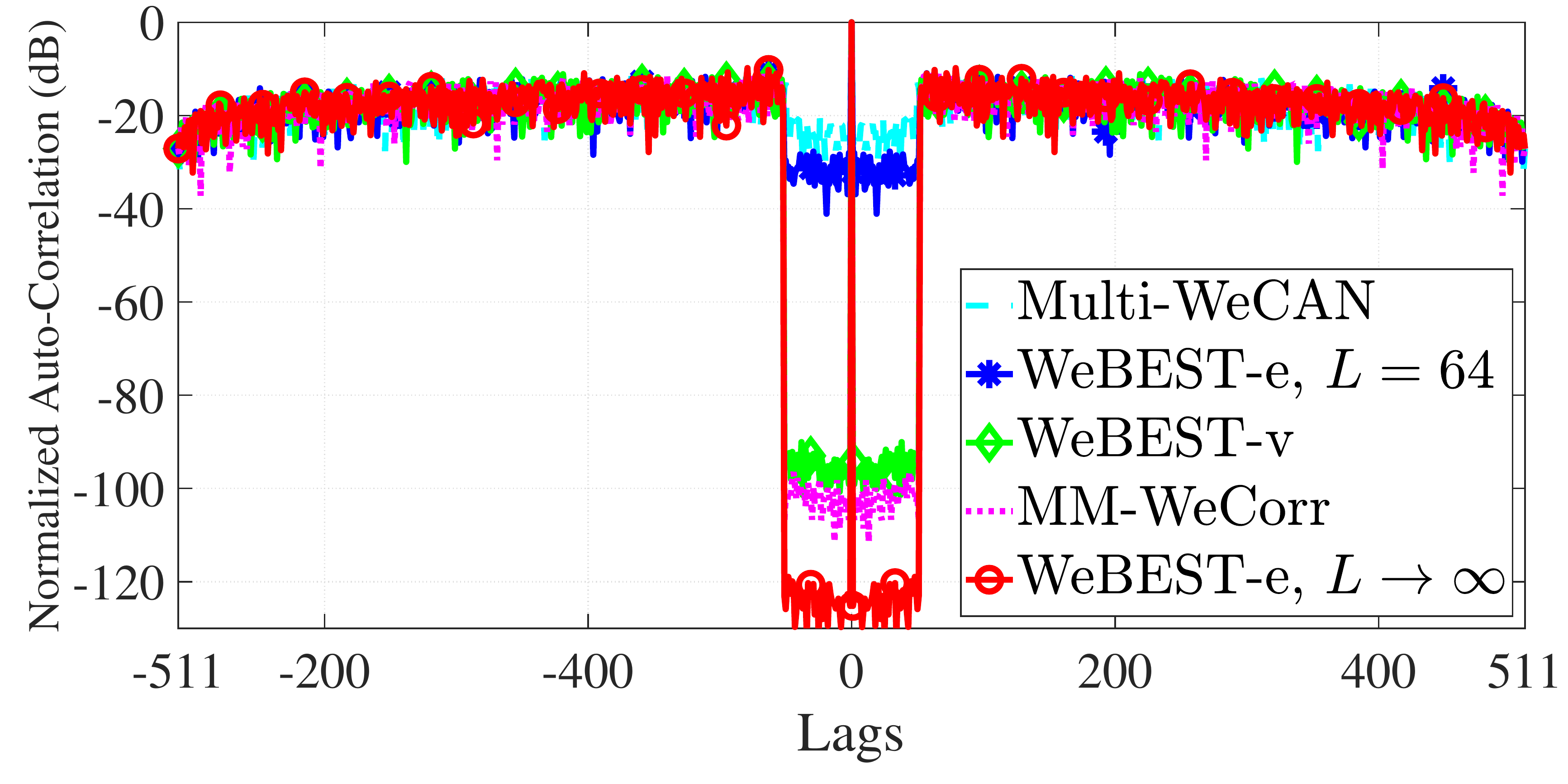}
		\caption[]{The Auto-Correlation of the second waveform.}\label{fig:AC2_vs_w}
    \end{subfigure}
    \caption[]{Comparison of the performance of the weighted \gls{ISL} minimization of the proposed method with \gls{MM}-WeCorr and Multi-WE\gls{CAN} unde discrete phase, entry and vector optimization ($p = 2$, $M=2$ and $N=512$).}\label{fig:ACCC_vs_w}
\end{figure*}

In \figurename{~\ref{fig:ACCC_vs_w}}, we compare the performance of the proposed method with \gls{MM}-WeCorr and Multi-We\gls{CAN} reported in \cite{7420715} and \cite{5072243} respectively. In this figure we assume that $p=2$, $M=2$ and $N=512$ and we consider to put nulls within range $\mathcal{V}=[-51, 51]$. As can be see, the proposed method outperforms the Multi-We\gls{CAN} method even by comparing the designed sequences with limited alphabet size.
The vector optimization approach has similar performance comparing to \gls{MM}-WeCorr. However, the entry optimization approach offers lower sidelobes in the lag region $\mathcal{V}=[-51, 51]$ when compared to \gls{MM}-WeCorr.

\subsection{Computational Time}
\begin{figure}
	\centering
	\includegraphics[width=1.0\columnwidth]{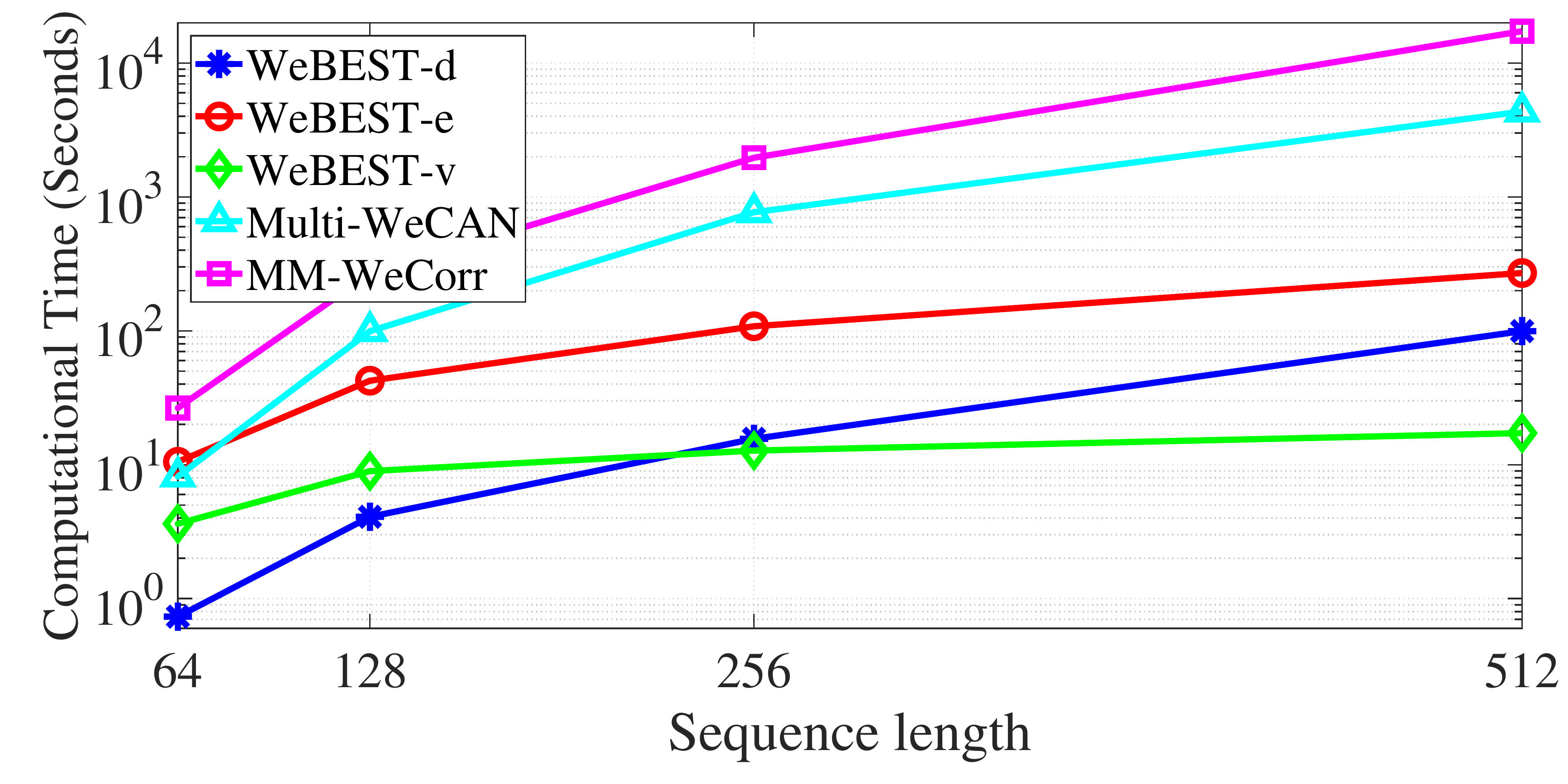}
	\caption{Comparison of the computational time of WeBEST with other methods. ($M=2$ and $L=64$)}
	\label{fig:CPUTime_vs_CANMM}
\end{figure}

In this subsection, we assess the computational time of \gls{WeBEST} and compare it with Multi-We\gls{CAN} and \gls{MM}-WeCorr. In this regard, we report the computational time by a desktop PC with Intel (R) Core (TM) i9-9900K CPU @ 3.60GHz with installed memory (RAM) 64.00 GB. \figurename{~\ref{fig:CPUTime_vs_CANMM}} shows the computational time of \gls{WeBEST}, Multi-We\gls{CAN} and \gls{MM}-WeCorr with $M=2$, $l=64$ and different sequence length. In this figure we assume that the desired lags are located at $\mathcal{V}=[-\lfloor 0.1N \rceil, \lfloor 0.1N \rceil]$. For fair comparison, we assume $\Delta \bX = 10^{-3}$ as stopping criteria for all methods.
Since in \gls{WeBEST}-v, we optimize a vector in each step, it is faster compare to other methods, especially with long sequence length. However, due to efficient formulation for entry-based optimization, \gls{WeBEST}-e has lower computational time when compare to Multi-We\gls{CAN} and \gls{MM}-WeCorr.

\section{Conclusion}
In this paper,  we considered the $\ell_p$-norm of auto- and cross-correlation functions of a set of sequences as the objective function and optimized the sequences under unimodular constraint using \gls{BSUM} framework. This problem formulation, provided further the flexibility for selecting $p$ and adapting waveforms based on the environmental conditions,  a key requirement for the emerging cognitive radar systems.
To tackle the problem, in every iterations of \gls{BSUM} algorithm, we utilized a local approximation function  to minimize the objective function. Specifically we introduced entry- and vector-based solutions where in the former we obtain critical points and  in the latter we obtain the gradient to find the optimized solution. We further used \gls{FFT}-based method for designing discrete phase sequences. 
Simulation results have illustrated the monotonicity of the proposed framework in minimizing the objective function. Besides, the proposed framework 
meets the lower bound in case of \gls{ISL} minimization, and outperform the counterparts in terms of \gls{PSL}, $l_0$-norm and computational time.
 
\appendices
\section{}\label{app:1}
The auto- and cross-correlation of $t^{th}$ transmitter can be written as $d^{th}$ entry as, \cite{8706639},
\par\noindent\small
\begin{equation} \label{eq:r_k wrt x_{t,d}}
\begin{aligned}
    &r_{t,t}(k) \triangleq \bar{c}_{ttdk} + \bar{a}_{ttdk}x_{t,d} + \bar{b}_{ttdk}x_{t,d}^*\\
    &r_{t,l}(k) \triangleq \bar{c}_{tldk} + \bar{a}_{tldk}x_{t,d}
\end{aligned}
\end{equation}
\normalsize
where,
\par\noindent\small
\begin{equation}
\begin{aligned}
	\bar{c}_{tldk} &\triangleq \textstyle \sum_{\substack{{n=1}\\{n \neq d}}}^{N-k}x_{t,n}x_{l,n+k}^*, \bar{a}_{tldk} \triangleq x_{l,d+k}^*I_A(d+k)\\
	\bar{c}_{ttdk} &\triangleq \textstyle \sum_{\substack{{n=1}\\{n \neq d, n \neq d-k}}}^{N-k}x_{t,n}x_{t,n+k}^* \\ \bar{a}_{ttdk} &\triangleq x_{t,d+k}^*I_A(d+k), \bar{b}_{ttdk} \triangleq x_{t,d-k}I_A(d-k)
\end{aligned}
\end{equation}
\normalsize
where, $I_A(p)$ is the indicator function of set $A = \left \{1, \dots, N\right \}$, i.e, $I_A(p) \triangleq \begin{dcases}
	1, & p \in A\\
	0, & p \notin A
	\end{dcases}$. 
Please note that the coefficients $c_{tldk}$ and $c_{ttdk}$ are depend on $\mathbf{x}_{t,-d}$ while $a_{tldk}$, $a_{ttdk}$ and $b_{ttdk}$ are depend on $x_{t,d}$. 

Therefore the weighted auto- and cross-correlation of $t^{th}$ transmitter becomes,
\par\noindent\small
\begin{equation} \label{eq:wr_k wrt x_{t,d}}
\begin{aligned}
    &w_k r_{t,t}(k) = c_{ttdk} + a_{ttdk}x_{t,d} + b_{ttdk}x_{t,d}^*\\
    &w_k r_{t,l}(k) = c_{tldk} + a_{tldk}x_{t,d}
\end{aligned}
\end{equation}
\normalsize
where
\par\noindent\small
\begin{equation}
\begin{aligned}
	a_{ttdk} &\triangleq w_k\bar{a}_{ttdk}, \ b_{ttdk} \triangleq w_k\bar{b}_{ttdk}, \ c_{ttdk} \triangleq w_k\bar{c}_{ttdk}, \\
	a_{tldk} &\triangleq w_k\bar{a}_{tldk}, \ c_{tldk} \triangleq w_k\bar{c}_{tldk}, 
\end{aligned}
\end{equation}
\normalsize

Substituting \eqref{eq:wr_k wrt x_{t,d}} in \tablename{~\ref{tab:w.r.t vector}}, the auto- and cross- terms of $f(\bX)$, $v_h^{\epsilon}(\bX)$ and $u(\bX)$ with respect to $x_{t,d}$ can be written as \tablename{~\ref{tab:w.r.t entry}}.

\section{}\label{app:2}
By substituting $x_{t,d} = e^{j\phi}$ in $v_{au,h}^{\epsilon}(\mathbf{x}_t)$ and $v_{cr,h}^{\epsilon}(\mathbf{x}_t)$ in \tablename{~\ref{tab:w.r.t entry}}, they can be expressed with respect to variable $\phi$. 
By expanding the absolute term and separating the $e^{jn\phi}$ terms by some mathematical manipulations, the auto- and cross-correlation term of $v_{h}^{\epsilon}(\phi)$ can be written as,
\par\noindent\small
\begin{equation}
v_{h,au}^{\epsilon}(\phi) = \textstyle \sum_{n=-2}^{2} \bar{v}_{h,n}e^{jn\phi}, \ v_{h,cr}^{\epsilon}(\phi) = \sum_{n=-1}^{1} \tilde{v}_{h,n}e^{jn\phi},
\end{equation}
\normalsize
where,
\par\noindent\small
\begin{fleqn}
\begin{equation*}
	\bar{v}_{h,-2} \triangleq \textstyle \sum_{k=-N+1}^{N-1} \gamma_{httk}(a_{ttdk}^*b_{ttdk}), \quad \bar{v}_{h,2} \triangleq \bar{v}_{h,-2}^*,
\end{equation*}
\end{fleqn}
\begin{fleqn}
\begin{equation*}
	\bar{v}_{h,-1} \triangleq \textstyle \sum_{k=-N+1}^{N-1}\gamma_{httk}(a_{ttdk}^* c_{ttdk} + c_{ttdk}^*b_{ttdk}), \ \bar{v}_{h,1} \triangleq \bar{v}_{h,-1}^*,
\end{equation*}
\end{fleqn}
\begin{fleqn}
\begin{equation*}
\bar{v}_{h,0} \triangleq \textstyle \sum_{k=-N+1}^{N-1}(\gamma_{httk}(|c_{ttdk}|^2 + |a_{ttdk}|^2 + |b_{ttdk}|^2) + \mu_{httk}). 
\end{equation*}
\end{fleqn}
\begin{fleqn}
\begin{equation*}
	\tilde{v}_{h,-1} \triangleq \textstyle 2\sum_{\substack{{l=1}\\{l \neq t}}}^{M} \sum_{k=-N+1}^{N-1}\gamma_{htlk}a_{tldk}^*c_{tldk}, \quad \tilde{v}_{h,1} \triangleq \tilde{v}_{h,-1}^*,
\end{equation*}
\end{fleqn}
\begin{fleqn}
\begin{equation*}
	\tilde{v}_{h,0} \triangleq \textstyle 2\sum_{\substack{{l=1}\\{l \neq t}}}^{M} \sum_{k=-N+1}^{N-1}(\gamma_{htlk}(|c_{tldk}|^2 + |a_{tldk}|^2) + \mu_{htlk}).
\end{equation*}
\end{fleqn}
\normalsize

Since, $v_{h}^{\epsilon}(\phi) = v_{h,au}^{\epsilon}(\phi) + v_{h,cr}^{\epsilon}(\phi) + v_{h,m}^{\epsilon}$, it can be written as \eqref{eq:Coeff_ui_vi}, where, 
\par\noindent\small
\begin{equation}
\begin{aligned}
&v_{h,-2} \triangleq \bar{v}_{h,-2}, \ v_{h,-1} \triangleq \bar{v}_{h,-1} + \tilde{v}_{h,-1} \\ &v_{h,0} \triangleq \bar{v}_{h,0} + \tilde{v}_{h,0} + v_{h,m}^{\epsilon}, \ v_{h,1} \triangleq \bar{v}_{h,-1}^* \ v_{h,2} \triangleq \bar{v}_{h,-2}^*
\end{aligned}
\end{equation}
\normalsize

Like wise, by substituting $x_{t,d} = e^{j\phi}$ in $u_{au}(\mathbf{x}_t)$ and $u_{cr}(\mathbf{x}_t)$, they can be expressed with respect to variable $\phi$.
Let, $\phi^{(i)} = \angle{x_{t,d}^{(i)}}$, hence, $w_kr_{t,l}^{(i)}(k) = c_{tldk} + a_{tldk}e^{j\phi^{(i)}}$ and $w_kr_{t,t}^{(i)}(k) = c_{ttdk} + a_{ttdk}e^{j\phi^{(i)}} + b_{ttdk}e^{-j\phi^{(i)}}$ respectively. Therefore $u_{au}(\phi)$ and $u_{cr}(\phi)$ becomes,
\par\noindent\small
\begin{equation*}
	u_{au}(\phi) = \textstyle \sum_{n=-2}^{2} \bar{u}_n e^{jn\phi} + \Re\left\{\sum_{n=-1}^{1} \hat{u}_n e^{jn\phi}\right\},
\end{equation*}
\normalsize
\par\noindent\small
\begin{equation*}
	u_{cr}(\phi) = \textstyle \sum_{n=-1}^{1} \tilde{u}_n e^{jn\phi} + \Re\left\{\sum_{n=-1}^{0} \check{u}_n e^{jn\phi}\right\},
\end{equation*}
\normalsize
Defining $\psi_{ttdk}' \triangleq \frac{\psi_{ttdk}}{|w_kr_{t,t}^{(i)}(k)|}$ and $\psi_{tldk}' \triangleq \frac{\psi_{tldk}}{|w_kr_{t,l}^{(i)}(k)|}$, it can be shown that, 
\par\noindent\small
\begin{fleqn}
\begin{equation*}
	\bar{u}_{-2} \triangleq \textstyle \sum_{k=-N+1}^{N-1} \eta_{ttdk}a_{ttdk}^*b_{ttdk}, \quad \bar{u}_{2} \triangleq \bar{u}_{-2}^*,
\end{equation*}
\end{fleqn}
\begin{fleqn}
\begin{equation*}
	\bar{u}_{-1} \triangleq \textstyle \sum_{k=-N+1}^{N-1} \eta_{ttdk}(a_{ttdk}^*c_{ttdk} + c_{ttdk}^*b_{ttdk}), \quad \bar{u}_{1} \triangleq \bar{u}_{-1}^*,
\end{equation*}
\end{fleqn}
\begin{fleqn}
\begin{equation*}
\bar{u}_{0} \triangleq \textstyle \sum_{k=-N+1}^{N-1}(\eta_{ttdk}(|c_{ttdk}|^2 + |a_{ttdk}|^2 + |b_{ttdk}|^2) + \nu_{ttdk}) 
\end{equation*}
\end{fleqn}
\begin{fleqn}
\begin{equation*}
	\hat{u}_{-1} \textstyle \triangleq \sum_{k=-N+1}^{N-1} \psi_{ttdk}'(|c_{ttdk}|^2 + c_{ttdk}^*a_{ttdk}e^{j\phi^{(i)}} + c_{ttdk}^*b_{ttdk}e^{-j\phi^{(i)}})
\end{equation*}
\end{fleqn}
\begin{fleqn}
\begin{equation*}
	\hat{u}_{0} \triangleq \textstyle \sum_{k=-N+1}^{N-1} \psi_{ttdk}'(|b_{ttdk}|^2e^{-j\phi^{(i)}} + b_{ttdk}^*a_{ttdk}e^{j\phi^{(i)}} + b_{ttdk}^*c_{ttdk})
\end{equation*}
\end{fleqn}
\begin{fleqn}
\begin{equation*}
	\hat{u}_{1} \triangleq \textstyle \sum_{k=-N+1}^{N-1} \psi_{ttdk}'(|a_{ttdk}|^2e^{j\phi^{(i)}} + a_{ttdk}^*b_{ttdk}e^{-j\phi^{(i)}} + a_{ttdk}^*c_{ttdk})
\end{equation*}
\end{fleqn}
\begin{fleqn}
\begin{equation*}
	\tilde{u}_{-1} \triangleq \textstyle 2\sum_{\substack{{l=1}\\{l \neq t}}}^{M} \sum_{k=-N+1}^{N-1}\eta_{tldk}c_{tldk}a_{tldk}^*, \quad \tilde{u}_{1} \triangleq \tilde{u}_{-1}^*
\end{equation*}
\end{fleqn}
\begin{fleqn}
\begin{equation*}
	\tilde{u}_{0} \triangleq \textstyle 2\sum_{\substack{{l=1}\\{l \neq t}}}^{M} \sum_{k=-N+1}^{N-1} (\eta_{tldk}(|c_{tldk}|^2 + |a_{tldk}|^2) + \nu_{tldk}))
\end{equation*}
\end{fleqn}
\begin{fleqn}
\begin{equation*}
	\check{u}_{-1} \triangleq \textstyle 2\sum_{\substack{{l=1}\\{l \neq t}}}^{M} \sum_{k=-N+1}^{N-1} \psi_{tldk}'(c_{tldk}a_{tldk}^* + |a_{tldk}|^2e^{j\phi^{(i)}})
\end{equation*}
\end{fleqn}
\begin{fleqn}
\begin{equation*}
	\check{u}_{0} \triangleq \textstyle 2\sum_{\substack{{l=1}\\{l \neq t}}}^{M} \sum_{k=-N+1}^{N-1}\psi_{tldk}'(|c_{tldk}|^2 + c_{tldk}^*a_{tldk}e^{j\phi^{(i)}})
\end{equation*}
\end{fleqn}
\normalsize

Since $u(\phi)$ is a real function, it can be written as $u(\phi) = \Re\left\{u_{au}(\phi) + u_{cr}(\phi) + u_m\right\}$, specifically can be written as \eqref{eq:Coeff_ui_vi}, where, 
\par\noindent\small
\begin{equation*}
\begin{aligned}
&u_{-2} \triangleq \bar{u}_{-2}, \quad u_{-1} \triangleq \tilde{u}_{-1} + \bar{u}_{-1} + \hat{u}_{-1} + \check{u}_{-1}\\
&u_{0} \triangleq \bar{u}_{0} + \hat{u}_{0} + \tilde{u}_{0} + \check{u}_{0} + u_m, \ u_{1} = \tilde{u}_{1} + \bar{u}_{1} + \hat{u}_{1}, \ u_{2} \triangleq \bar{u}_{2}
\end{aligned}
\end{equation*}
\normalsize

\section{}\label{app:3}
Substituting $e^{jn\phi} = \cos{(n\phi)} + j\sin{(n\phi)}$ in $u'(\phi)$ and separating the real and imaginary part, $u'(\phi)$ becomes, 
\par\noindent\small
\begin{equation}
\begin{aligned}
u'(\phi) &= \xi_0\cos^2(\phi) + \xi_1\sin^2(\phi) + \xi_2\sin(\phi)\cos(\phi) \\
         &+ \xi_3\cos(\phi) + \xi_4\sin(\phi)
\end{aligned}
\end{equation}
\normalsize
where, $\xi_0 \triangleq 2\Im\{u_{-2} - u_2\}$, $\xi_1 \triangleq 2\Im\{u_2 - u_{-2}\}$, $\xi_2 \triangleq - 4\Re\{u_2 + u_{-2}$\}, $\xi_3 \triangleq \Im\{u_{-1} - u_1\}$ and $\xi_4 \triangleq -\Re\{u_{-1} + u_1\}$. Using the change variable $z \triangleq \tan(\frac{\phi}{2})$ and substituting $\cos(\phi) = {(1-z^2)}/{(1+z^2)}$, $\sin(\phi) = {2z}/{(1+z^2)}$ in $u'(\phi)$, it can be written as, $u'(z)=\frac{\sum_{k=0}^{4} s_kz^k}{(1+z^2)^2}$, where, 
\par\noindent\small
\begin{equation}
\begin{aligned}
&s_0 \triangleq \xi_0 + \xi_3, \ s_1 \triangleq 2(\xi_2 + \xi_4), \ s_2 \triangleq  2(2\xi_1 - \xi_0), \\ 
&s_3 \triangleq 2(\xi_4 - \xi_2), \ s_4 \triangleq \xi_0 - \xi_3
\end{aligned}
\end{equation}
\normalsize

Likewise, considering $z \triangleq \tan(\frac{\phi}{2})$, the roots of $v_h^{\epsilon'}(\phi)$ can be equivalently obtained by solving $\frac{\sum_{k=0}^{4} q_{h,k}z^k}{(1+z^2)^2}=0$, where,
\par\noindent\small
\begin{equation}
\begin{aligned}
&q_{h,0} \triangleq \varkappa_0 + \varkappa_3, \ q_{h,1} \triangleq 2(\varkappa_2 + \varkappa_4), \ q_{h,2} \triangleq  2(2\varkappa_1 - \varkappa_0), \\ 
&q_{h,3} \triangleq 2(\varkappa_4 - \varkappa_2), \ q_{h,4} \triangleq \varkappa_0 - \varkappa_3,
\end{aligned}
\end{equation}
\normalsize
and, $\varkappa_0 \triangleq -4\Im\{v_{h,2}\}$, $\varkappa_1 \triangleq 4\Im\{v_{h,2}\}$, $\varkappa_2 \triangleq - 8\Re\{v_{h,2}\}$, $\varkappa_3 \triangleq -2\Im\{v_{h,1}\}$ and $\varkappa_4 \triangleq -2\Re\{v_{h,1}\}$. 


\section{}\label{app:4}
Under discrete phase constraint, since the phases are chosen from finite alphabet ($\phi \in \Omega_L$) the objective function can be written with respect to the the indices of $\Omega_L$ as follows,
\par\noindent\small
\begin{equation}\label{eq:f_l}
\begin{aligned}
    &f(l') = f(\bX_{-t}) + \textstyle 2\sum_{\substack{{l=1}\\{l \neq t}}}^{M}\sum_{k=-N+1}^{N-1}|a_{tldk} + c_{tldk}e^{-j2\pi\frac{l'-1}{L}}|^p \\
    &+\sum_{k=-N+1}^{N-1} |a_{ttdk} + c_{ttdk}e^{-j2\pi\frac{l'-1}{L}} + b_{ttdk}e^{-j4\pi\frac{l'-1}{L}}|^p
\end{aligned}
\end{equation}
\normalsize
Observe that $a_{tldk} + c_{tldk}e^{-j2\pi\frac{l'-1}{L}}$ and $a_{ttdk} + c_{ttdk}e^{-j2\pi\frac{l'-1}{L}} + b_{ttdk}e^{-j4\pi\frac{l'-1}{L}}$ exactly follow the definition of $L$-points \gls{DFT} of sequences $\{a_{tldk}, c_{tldk}\}$ and $\{a_{ttdk}, c_{ttdk}, b_{ttdk}\}$ respectively. Therefore, $f(l')$ can be written as \eqref{eq:P_l}.

\ifCLASSOPTIONcaptionsoff
  \newpage
\fi



%




\bibliographystyle{IEEEtran}
\bibliography{IEEEabrv,J2_TSP_2021_Arxiv}

%








\end{document}